\documentclass[12pt]{iopart}
\usepackage{graphics,graphicx,subfigure}

\begin{document}

\title{Tricriticality for dimeric Coulomb molecular crystals in ground state}

\author{Igor Trav\v{e}nec and Ladislav \v{S}amaj}

\address{Institute of Physics, Slovak Academy of Sciences, 
D\'ubravsk\'a cesta 9, 84511 Bratislava, Slovakia}
\ead{fyzitrav@savba.sk}
\vspace{10pt}
\begin{indented}
\item[]
\end{indented}

\begin{abstract}
We study the ground-state properties of a system of dimers. 
Each dimer consists in a pair of equivalent charges at a fixed distance,
immersed in a neutralizing homogeneous background. 
All charges interact pairwisely by Coulomb potential.
The dimer centers form a two-dimensional rectangular lattice with the aspect
ratio $\alpha\in [0,1]$ and each dimer is allowed to rotate around its center. 
The previous numerical simulations, made for the more general Yukawa
interaction, indicate that only two basic dimer configurations can appear:
either all dimers are parallel or they have two different angle orientations 
within alternating (checkerboard) sublattices.
As the dimer size increases, two second-order phase transitions,
related to two kinds of the symmetry breaking in dimer's orientations,
were reported.
In this paper, we use a recent analytic method based on an expansion of 
the interaction energy in Misra functions which converges quickly and 
provides an analytic derivation of the critical behaviour.
Our main result is that there exists a specific aspect ratio of 
the rectangular lattice $\alpha^*=0.71410684000071\ldots$ which divides
the space of model's phases onto two distinct regions. 
If the lattice aspect ratio $\alpha>\alpha^*$, we recover both types of 
the second-order phase transitions and find that they are of mean-field type 
with the critical exponent $\beta = 1/2$. 
If $\alpha<\alpha^*$, the phase transition associated with the discontinuity 
of dimer's angles on alternating sublattices becomes of first order.
For $\alpha=\alpha^*$, the first- and second-order phase transitions meet 
at the tricritical point, characterized by the different critical index 
$\beta = 1/4$.
Such phenomenon is known from literature about the Landau theory of 
one-component fields, but in our two-component version the scenario is 
more complicated: the component which is already in the symmetry-broken state
at the tricritical point also interferes and exhibits unexpectedly 
the mean-field singular behaviour.
\end{abstract}


\vspace{2pc}

\submitto{\JPA}

\maketitle

\section{Introduction} \label{Sec1}
Two-dimensional (2D) systems of charged colloidal particles in periodic 
external potentials have been investigated both experimentally and 
theoretically. 
The external field can be generated by optical trapping methods \cite{Bleil06},
or by producing technologically demanding monolayers on a substrate 
\cite{Rycenga09}.
Another experiment \cite{Ma12}, when an electric field is applied to colloidal 
dimers, leads to 2D dimeric crystals. 
The possibility of experimental realization of certain systems boosted 
theoretical studies of low-dimensional colloidal systems.

In this paper, we restrict ourselves to the study of the ground state of 
2D dimeric structures at zero temperature.
It is obvious that anisotropic particles can create a wider range of 
distinct phases than the spherical ones.
Such particles can be represented by anisotropic colloids themselves, 
but they can be effectively created as complexes of identical two (dimer) 
\cite{Mikulis04,ElShawish08,Trizac10,Agra04,Reichhardt02} or three (trimer) 
\cite{Mikulis04,ElShawish08,Reichhardt02,Brunner02} bound particles 
per potential minimum of the external potential. 
Even the case of four particles was studied previously \cite{Reichhardt02}.
Another problem is the ground-state of dipoles \cite{Trizac10}, i.e., 
units with two oppositely charged particles, or clusters of two negative 
and one positive charges \cite{ElShawish11}.
The filling can be even rational \cite{Reichhardt03,Reichhardt05}.
The model complexity is usually reduced by considering $n$-mers
as rigid entities fixed on a lattice structure (given implicitly by
the confining potential), with an orientational degree of freedom only. 
A more complete description is provided within the flexible model in which 
the center of an $n$-mer can move from the lattice position \cite{ElShawish08}
during the energy minimization.
The application of additional external fields is also possible \cite{Rycenga09}.

As concerns the type of pair interactions, the mostly considered interaction 
of colloids at distance $r$ is the Yukawa one $\propto \rme^{-\kappa r}/r$, 
i.e., the screened Coulomb interaction \cite{Dobnikar02}.
If the inverse screening length $\kappa\to 0$, we get the pure Coulomb 
interaction which was used, e.g., in dimeric bilayers \cite{Lobaskin07}.
In the case of small dipoles, the potential $\propto 1/r^3$ is considered
\cite{Granz16}.
Besides the ground state at $T=0$, the phase behaviour at nonzero temperatures 
$T$, up to the melting point, was studied as well \cite{ElShawish12,Sarlah05}.

We shall concentrate on molecular crystals formed by rigid dimers with 
Coulomb interaction whose centers are rigidly pinned to the sites of 
a 2D rectangular lattice with the aspect ratio $\alpha\in [0,1]$. 
Each dimer is allowed to rotate around its center and we study their 
orientational ordering phase transitions as the dimer size increases. 
Such systems, with a more general Yukawa interaction of particles, were 
investigated experimentally \cite{Ma12,Brunner02} and by numerical simulations 
combined with analytic considerations \cite{Agra04,Sarlah05,Sarlah07}.
Numerical simulations indicate that only two basic dimer configurations can 
appear: either all dimers are parallel or they have two different angle 
orientations within alternating (checkerboard) sublattices 
\cite{Trizac10,ElShawish12}.
Increasing successively the dimer distance, two second-order phase transitions,
related to two kinds of the symmetry breaking in dimer's orientations,
were reported.

In this paper, we use a recent method for calculating the energy lattice 
summations, applied originally to Coulomb \cite{Samaj12} and Yukawa 
\cite{Travenec15} bilayers.
The method is based on a series of transformations with Jacobi theta
functions which permits one to write the interaction energy as a quickly 
converging series of generalized Misra functions \cite{Misra40}. 
Misra functions can be expanded in powers of the order parameter near 
the critical point, leading to an exact Landau form of the ground-state 
energy function.
The type of the critical behaviour is thus identified and the critical 
point can be specified with an arbitrary accuracy. 

Our main result is that there exists a specific aspect ratio of 
the rectangular lattice $\alpha^*$ which divides the space of model's phases 
onto two distinct regions. 
If $\alpha>\alpha^*$, we recover both previously observed second-order 
phase transitions and find that they are of mean-field type with the critical 
exponent $\beta = 1/2$. 
If $\alpha<\alpha^*$, the phase transition associated with the discontinuity 
of dimer's angles on alternating sublattices becomes of first order.
The first- and second-order phase transitions meet at the tricritical point 
($\alpha=\alpha^*$), characterized by the different critical index 
$\beta = 1/4$.
The atypical value of $\beta$ is known in the Landau theory of one-component 
fields, see e.g. book \cite{Chaikin}. 
But in our two-component case the scenario is more complicated: the component 
which is already in the symmetry-broken state at the tricritical point also 
interferes and exhibits unexpectedly the singular behaviour of mean-field type.

The article is organized as follows.
In section \ref{Sec2} we introduce the model and review its possible phases.
We show typical phase diagrams, one for $\alpha>\alpha^*$ with two second-order
phase transitions and the other for $\alpha<\alpha^*$ with one second- and 
one first-order phase transitions, and identify the respective order 
parameters.
Analytic formulas for the interaction energy and their expansions around 
relevant points are presented in section \ref{Sec3}. 
Section \ref{Sec4} deals with the first type of second-order phase transition
between two (disordered and ordered) homogeneous phases with all dimers being 
parallel; the square lattice is treated as a special case.
The transition between the phase with all dimers being parallel and the other 
with two alternating non-equivalent sublattices is studied in the subsequent
three sections.
Section \ref{Sec5} deals with the second-order phase transition for 
$\alpha>\alpha^*$, section \ref{Sec6} is devoted to to the study 
of tricriticality at $\alpha=\alpha^*$ and section \ref{Sec7} discusses 
the first-order phase transition for $\alpha<\alpha^*$.
Section \ref{Sec8} is a short recapitulation with concluding remarks.

\section{The model and its phases} \label{Sec2}
We consider dimers which consist of two equivalent pointlike particles 
with unit charge $q=1$ at distance $2d$. 
Dimer centers form a periodic 2D rectangular lattice $\Lambda$ of sides 
$a$ along the $x$-axis and $\alpha a$ along the $y$-axis,
with $\alpha$ being the aspect ratio.
There is a symmetry with respect to the rotation of the whole lattice by 
the right angle, $\alpha \leftrightarrow 1/\alpha$, so we can restrict 
ourselves to the interval $0<\alpha\le 1$ where $\alpha=1$ for 
the square lattice.
To simplify the notation, we shall work in length units of $a=1$.
The lattice $\Lambda$ consists of points
\begin{equation} \label{Lambda}
\Lambda = \left\{ i+\alpha j; i,j=0,\pm 1,\pm 2,\ldots \right\} .
\end{equation}
It can be decomposed onto two alternating sublattices,
$\Lambda = \Lambda_1 \cup \Lambda_2$, such that the sites of $\Lambda_1$ 
are those with $i+j={\rm even}$ and the sites of $\Lambda_2$ are those with 
$i+j={\rm odd}$.

Each dimer is allowed to rotate around its center and its orientation is 
given by the angle $\varphi$ with respect to the $x$-axis.
To let the dimer rotate freely without colliding with its nearest neighbour, 
the dimer length must be smaller than the shorter side of the rectangle, i.e., 
\begin{equation} \label{constraint}
d < \frac{\alpha}{2} .
\end{equation}
From between two free parameters $\alpha$ and $d$, we fix $\alpha$ and
change continuously $d\in (0,\alpha/2)$; the dependence on $\alpha$ will not
be indicated, except for specific cases to avoid misunderstanding.
The particles interact by the three-dimensional (3D) Coulomb $1/r$ potential
(in Gauss units and with the dielectric constant $\varepsilon=1$)
and are immersed in a homogeneous neutralizing background charge density
which cancels the divergencies from the lattice summation due
to the long-range Coulomb tail. 

Numerical simulations with Yukawa potential \cite{Trizac10,ElShawish12} reveal
that only two types of dimer configurations minimize the interaction energy.
The first type corresponds to the spatially homogeneous phases when all 
dimers are parallel with the same angle $\varphi$, i.e., the energy depends
only on one field component. 
There are obvious $\varphi\to -\varphi$ and $\varphi\to \varphi+\pi$
symmetries of the interaction energy; here, we choose $\varphi\in [0,\pi/2]$.
The disordered phase is characterized by $\varphi=0$ and the ordered phase 
by $\varphi>0$, with the angle $\varphi$ being the order parameter.
The second type of the ordered ground state is characterized by two distinct 
orientations $\varphi_1$ and $\varphi_2$ on the alternating sublattices 
$\Lambda_1$ and $\Lambda_2$, respectively.
We shall refer to such phases $(\varphi_1,\varphi_2)$ as the inhomogeneous 
ones, with two nonzero field components and the order parameter 
proportional to $\varphi_1-\varphi_2$.
Within the $(\varphi_1,\varphi_2)$ notation, the two homogeneous phases are 
identified by $(0,0)$ (disordered phase) and by $(\varphi,\varphi)$ 
with $\varphi\ne 0$ (ordered phase). 

\begin{figure}[]
\centering
\subfigure[]{
\label{f1a}
\includegraphics[clip,width=.47\linewidth]{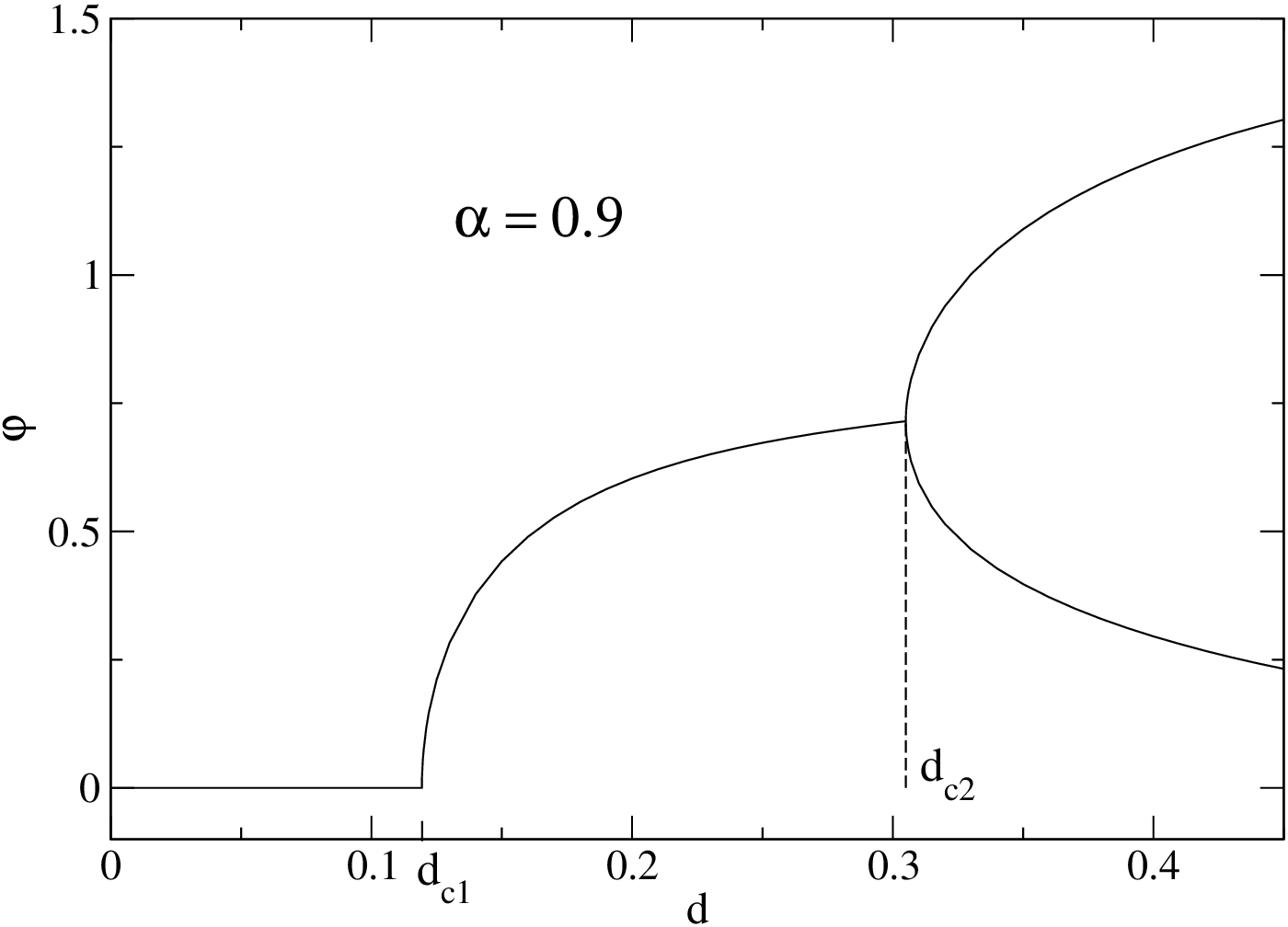}}
\quad
\centering
\subfigure[]{
\label{f1b}
\includegraphics[clip,width=.47\linewidth]{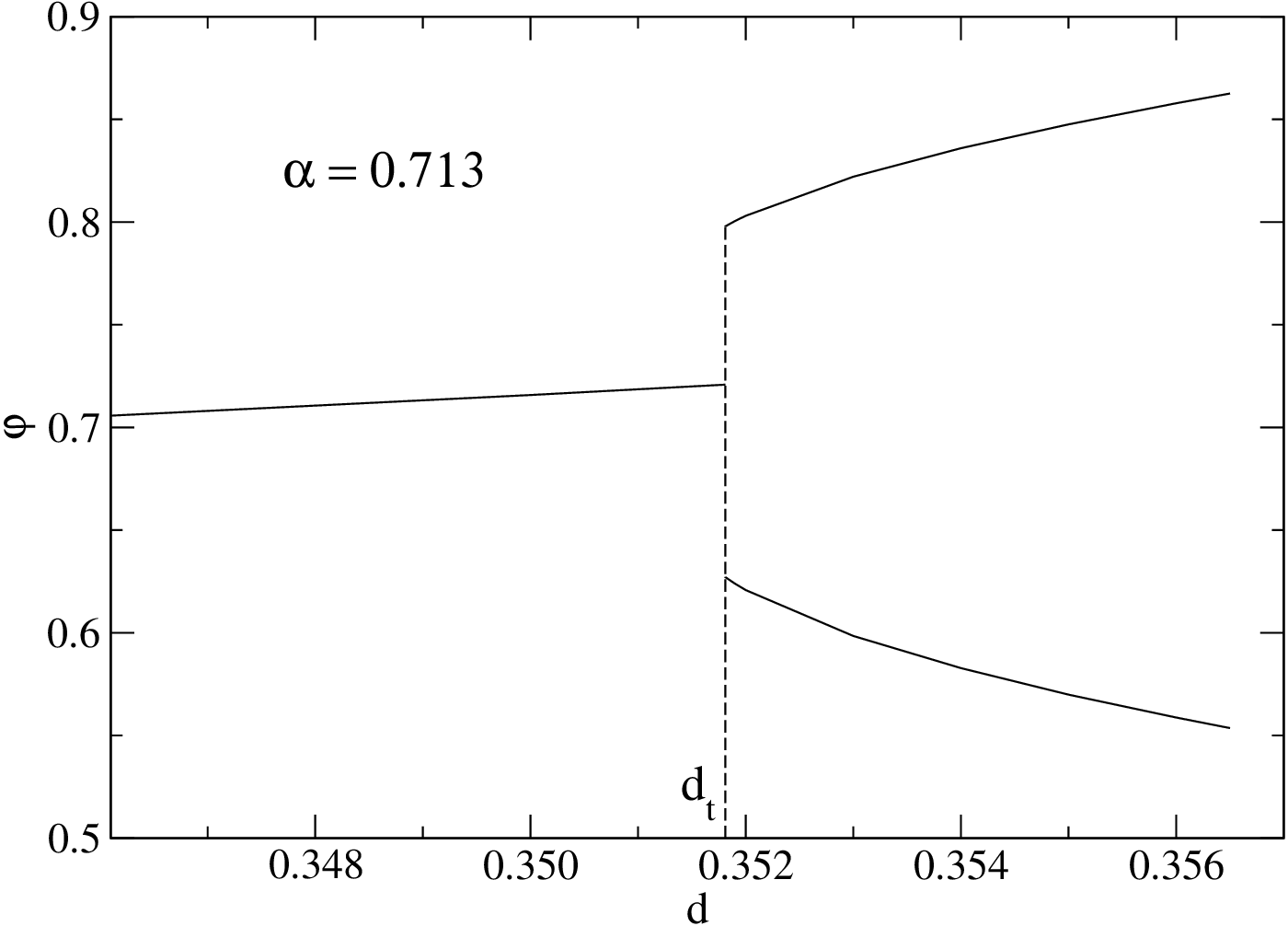}}
\caption {Angle(s) as function(s) of dimer's half-size $d$. 
(a) The aspect ration $\alpha=0.9$. 
Two critical points $d_{c1}$ and $d_{c2}$ separate the phases $(0,0)$ and
$(\varphi,\varphi)$, $(\varphi,\varphi)$ and $(\varphi_1,\varphi_2)$,
respectively.
(b) $\alpha=0.713$. The first-order transition point $d_{t}$ separates 
the phases $(\varphi,\varphi)$ and $(\varphi_1,\varphi_2)$, 
the critical point $d_{c1}$ is out of the plotted range.}
\end{figure}

For the Coulomb version of the model with a relatively large value of
the aspect ratio $\alpha=0.9$, the above scenario is represented 
graphically in Fig. \ref{f1a}.
The figure shows the dependence of the angles(s) on $d$ which we got 
in this work by minimizing the energy per particle.
One can see three phases: the homogeneous disordered phase $(0,0)$ for 
$0<d<d_{c1}$ ($d_{c1}=0.119354\ldots$), the homogeneous symmetry-broken phase 
$(\varphi,\varphi)$ for $d_{c1}<d<d_{c2}$ ($d_{c2}=0.3050058\ldots$) 
and the inhomogeneous phase $(\varphi_1,\varphi_2)$ 
for $d_{c2}<d<\alpha/2=0.45$.
Similar results were obtained in Ref. \cite{ElShawish08,Trizac10} for 
the same value of $\alpha$, but with non-zero $\kappa$.
Both phase transitions at $d_{c1}$ and $d_{c2}$ are of second order
and they are related to specific symmetries of the interaction energy
per particle $E(\varphi_1,\varphi_2,d)$. 
For the homogeneous phases, we shall use a simplified notation
\begin{equation}
E_0(\varphi,d) \equiv E(\varphi,\varphi,d) .
\end{equation}
For the transition between the homogeneous phases $(0,0)$ and 
$(\varphi,\varphi)$, the obvious mirror symmetry
\begin{eqnarray} \label{sym1}
E_0(\varphi,d) =  E_0(-\varphi,d)
\end{eqnarray}
is relevant.
For the transition between the homogeneous phase $(\varphi,\varphi)$ 
and the inhomogeneous one with alternating sublattices 
$(\varphi_1,\varphi_2)$, we make the transformation of 
the angle variables $\varphi_1=\varphi+\delta\varphi$ and
$\varphi_2=\varphi-\delta\varphi$, so that
\begin{equation}\label{phitransf}
\varphi=\frac{\varphi_1+\varphi_2}{2} , \qquad
\delta\varphi=\frac{\varphi_1-\varphi_2}{2} , \qquad
E(\varphi_1,\varphi_2,d) \equiv E(\varphi,\delta\varphi,d) . 
\end{equation}
If we interchange the sublattices $\Lambda_1$ and $\Lambda_2$, 
i. e. $\varphi_1\leftrightarrow \varphi_2$, the energy remains unchanged. 
This means the symmetry of the energy with respect to the transformation
$\delta\varphi\to -\delta\varphi$,
\begin{eqnarray} \label{sym2}
 E(\varphi,\delta\varphi,d) =  E(\varphi,-\delta\varphi,d) .
\end{eqnarray}
In both cases (\ref{sym1}) and (\ref{sym2}), the expansion of the energy 
around critical points involves only even powers of the order parameter, 
like in the general Landau theory of second-order phase transitions.

The numerical results for the smaller value of the aspect ratio 
$\alpha=0.713$ are presented in Fig. \ref{f1b}.
The first-type transition of Coulomb dimers between two homogeneous phases at 
$d_{c1}$ (not shown in the figure) is again of second-order.
On the other hand, our numerical results indicate that the second-type
transition from the homogeneous $(\varphi,\varphi)$ to the inhomogeneous 
$(\varphi_1,\varphi_2)$ phases becomes discontinuous (of first order).
The fact that this phenomenon was not observed in the previous simulations
of Yukawa crystals \cite{ElShawish08,Trizac10} is probably related to
the constraint (\ref{constraint}) which makes the corresponding region 
of model's parameters relatively small.
Our main task is to find the aspect-ratio value $\alpha^*$ at which
the second- and first-order transitions meet at the tricritical point.
We aim at describing fundamental changes in the critical behaviour at
the tricritical point.
To see them one has to approach with the aspect ratio of the rectangular 
lattice extremely close to the exact $\alpha^*$, which is practically 
impossible in numerical simulations.
On the other hand, our analytic approach permits to determine $\alpha^*$
with a high accuracy as the solution of a closed-form equation and to derive
exactly the form of the critical singularity at the tricritical point.    

\section{Analytic formulas for the energy} \label{Sec3}
In the most general case of alternating sublattices $(\varphi_1,\varphi_2)$,
the elementary cell is the rectangle of sides 2 and $2\alpha$.
To calculate the Coulomb energy per particle, we first average over 
positions of the reference particle on the two non-equivalent sublattices.
In both cases, the reference particle interacts with all particles inside
the elementary rectangle and with their periodic images.
In this way we get 
\begin{eqnarray} 
E(\varphi_1,\varphi_2,d) & = & \frac{1}{2}\Bigg\{ 
\Sigma_1\left[\alpha\right]
+ \Sigma_2\left[\textstyle{\alpha,\frac{1}{2},\frac{1}{2}}\right]
\nonumber \\ &&
+ \Sigma_2\left[\textstyle{\alpha,\frac{1}{2}
+ \frac{d}{2}\left(\cos{\varphi_1}-\cos{\varphi_2}\right),
\frac{d}{2\alpha}\left(\sin{\varphi_1}-\sin{\varphi_2}\right)}\right]
\nonumber \\ &&
+\Sigma_2\left[\textstyle{\alpha,\frac{1}{2}+\frac{d}{2}\left(\cos{\varphi_1}
+\cos{\varphi_2}\right),
\frac{d}{2\alpha}\left(\sin{\varphi_1}+\sin{\varphi_2}\right)}\right]
\nonumber \\ &&
+\Sigma_2\left[\textstyle{\alpha,
\frac{d}{2}\left(\cos{\varphi_1}-\cos{\varphi_2}\right),
\frac{1}{2}+\frac{d}{2\alpha}\left(\sin{\varphi_1}-\sin{\varphi_2}\right)}
\right] \nonumber \\ &&
+ \Sigma_2\left[\textstyle{\alpha,
\frac{d}{2}\left(\cos{\varphi_1}+\cos{\varphi_2}\right),
\frac{1}{2}+\frac{d}{2\alpha}\left(\sin{\varphi_1}+\sin{\varphi_2}\right)}
\right]\Bigg\} \nonumber \\ &&
+\frac{1}{4}\Bigg\{
\Sigma_2\left[\textstyle{\alpha,d\cos{\varphi_1},
\frac{d}{\alpha}\sin{\varphi_1}}\right] \nonumber
+\Sigma_2\left[\textstyle{\alpha,d\cos{\varphi_2},
\frac{d}{\alpha}\sin{\varphi_2}}\right] \nonumber \\ &&
+\Sigma_2\left[\textstyle{\alpha,\frac{1}{2}+d\cos{\varphi_1},
\frac{1}{2}+\frac{d}{\alpha}\sin{\varphi_1}}\right] \nonumber \\ &&
+\Sigma_2\left[\textstyle{\alpha,\frac{1}{2}+d\cos{\varphi_1},
\frac{1}{2}+\frac{d}{\alpha}\sin{\varphi_1}}\right]\Bigg\} . \label{en}
\end{eqnarray}
Here, the function $\Sigma_1(\alpha)$ sums the interaction energies of 
the reference particle, sitting on a given site of a rectangular lattice
with the aspect ratio $\alpha$, with all other particles on the remaining
sites of this rectangular structure \cite{Travenec16}:
\begin{eqnarray} \label{s1}
\Sigma_1(\alpha) & = & \sqrt{\pi}\left( 
\sum_{j,k=-\infty \atop \{j,k\}\ne\{0,0\}}^\infty
\frac{1}{\sqrt{j^2\alpha+k^2/\alpha}} + E_B^{(1)} \right) \nonumber \\ 
& = & 4 \sum_{j=1}^\infty \left[z_{3/2}(j^2\alpha) 
+ z_{3/2}\left(\textstyle{\frac{j^2}{\alpha}}\right) \right] + 8 
\sum_{j,k=1}^\infty z_{3/2}\left(\textstyle{j^2\alpha+\frac{k^2}{\alpha}}\right) 
- 4\sqrt{\pi},
\end{eqnarray}
where $E_B^{(1)}$ involves the energy of background-particle and 
background-background interactions and
\begin{equation} \label{Misra}
z_{\nu}(y) = \int_0^{1/\pi} \frac{{\rm d}t}{t^{\nu}} 
\exp\left( -\frac{y}{t}\right) , \qquad y>0 
\end{equation}
are the Misra functions \cite{Misra40}. 
The conversion of the lattice sum onto the series of Misra functions,
based on the Poisson summation formula and specific properties of
the Jacobi theta functions, is explained in 
Refs. \cite{Samaj12,Travenec15,Travenec16}.  
The representation of $z_{3/2}(y)$ in terms of the complementary error 
function is given in Eq. (\ref{znuy}).
The function $\Sigma_2(\alpha,a_1,a_2)$ sums the energy over all sites of 
the rectangular lattice, with the reference point shifted by the 
relative coordinates $a_1$ and $a_2$ from its nearest neighbour on the
rectangular structure:
\begin{eqnarray} \label{s2}
\Sigma_2(\alpha,a_1,a_2) & = & \sqrt{\pi} \left[\sum_{j,k=-\infty}^\infty 
\frac{1}{\sqrt{\frac{(a_1+j)^2}{\alpha}+(a_2+k)^2\alpha}}
+ E_B^{(2)} \right] \nonumber \\ & = & 2\sum_{j=1}^{\infty} 
\left[\cos(2\pi j a_1) z_{3/2}(j^2\alpha)
+\cos(2\pi j a_2)z_{3/2}\left(\textstyle{\frac{j^2}{\alpha}}\right)\right] 
\nonumber\\ & & + 4\sum_{j,k=1}^\infty \cos(2\pi j a_1)\cos(2\pi k a_2)
z_{3/2}\left(\textstyle{j^2\alpha+\frac{k^2}{\alpha}}\right) \nonumber\\
& & + \sum_{j,k=-\infty}^\infty 
z_{3/2}\left[\textstyle{\frac{(j+a_1)^2}{\alpha}+(k+a_2)^2\alpha}\right] 
- 2\sqrt{\pi} .
\end{eqnarray}
The series in Misra functions (\ref{s1}) and (\ref{s2}) are quickly
converging; the previous calculations \cite{Samaj12,Travenec15,Travenec16}
show that the truncation of the series over $j,k$ at $M=4$ ensures 
precision of 17 decimal digits. 
This precision is not sufficient close to the tricritical point,
so in this work we truncate the series at $M=6$ to ensure more 
than 25 decimal digits precision.
Such need of accuracy is connected with the fact that if one wants to get 
the values of $\varphi_1$ and $\varphi_2$ with $N$-digits precision, 
the energy minimization has to be done with roughly $2N$-digits accuracy.
The computation of one energy value by our Misra series requires the CPU
time of the order of one second on the standard PC.

In the homogeneous case $\varphi_1=\varphi_2=\varphi$, the energy formula 
(\ref{en}) simplifies itself substantially and takes the form
\begin{equation} \label{e0}
E_0(\varphi,d) = \Sigma_1(\alpha) + \Sigma_2\left(
\textstyle{\alpha,2d\cos{\varphi},\frac{2d}{\alpha}\sin{\varphi}}\right) .
\end{equation}
In the vicinity of the first-type phase transition at $d_{c1}$, 
due to the symmetry (\ref{sym1}) this energy can be expanded in even powers 
of small $\varphi$. 
Using the expansion formula (\ref{zexp}) for the Misra functions, 
the expansion of the energy reads as
\begin{eqnarray} \label{e0ex}
E_0(\varphi,d) = E_0(0;d) + g_2(d) \varphi^2 + g_4(d) \varphi^4 +
{\cal O}(\varphi^6) ,
\end{eqnarray}
where the explicit formulas for the functions $g_2$ and $g_4$ are given 
in Eqs. (\ref{g2}) and (\ref{g4}), respectively.

Analogously, with respect to the symmetry (\ref{sym2}), the most general 
energy (\ref{en}) can be expanded in even powers of $\delta\varphi$ as follows
\begin{equation} \label{enex}
E(\varphi,\delta\varphi,d) = E_0(\varphi,d) + h_2(\varphi,d)\delta\varphi^2 
+ h_4(\varphi,d)\delta\varphi^4 + {\cal O}(\delta\varphi^6).
\end{equation}
The absolute term is simply the homogeneous ($\delta\varphi=0$) energy
(\ref{e0}), the function $h_2$ is presented in Eq. (\ref{h2}) and 
$h_4$ is too lengthy to be written explicitly.

\section{Second-order transition between homogeneous phases} \label{Sec4}
This section concerns the homogeneous phases with all dimers being parallel. 
We first study the limiting case $\alpha\to 1$ (close to the square lattice) 
and then describe the second-order transition between the disordered $(0,0)$ 
and symmetry-broken $(\varphi,\varphi)$ ($\varphi\ne 0$) phases.

\subsection{The limit of the square lattice, $\alpha\to 1$}
\begin{figure}[]
\begin{center}
\includegraphics[clip,width=0.6\textwidth]{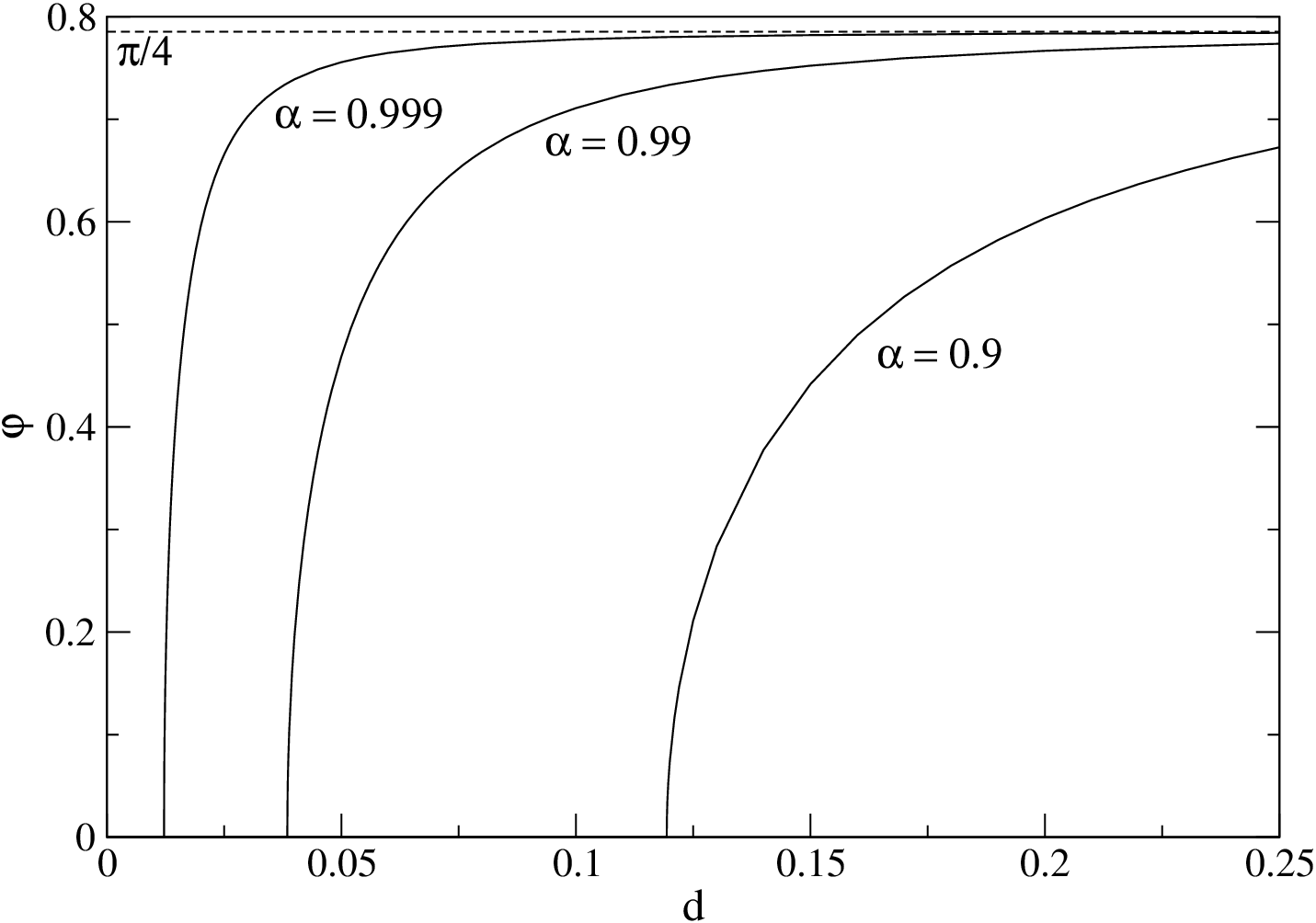}
\caption{The dimeric angle $\varphi$ as the function of $d$ for several 
values of the aspect ration $\alpha$. 
The case of the square lattice ($\alpha=1$) corresponds to the dashed line 
with $\varphi=\pi/4$.}
\label{fig:phida1}
\end{center}
\end{figure}

In the case of the square lattice with $\alpha=1$ it was found for Yukawa 
interactions \cite{ElShawish08,Trizac10} that for small values of $d$ 
the inhomogeneous phase $(0,\pi/2)$ appears, followed (via a first-order 
transition) by the homogeneous phase $(\pi/4,\pi/4)$ which is dominant
up to $d_{c2}$.
Numerically, we have minimized the general energy (\ref{en}) with 
the possibility of $\varphi_1\ne\varphi_2$.
We observed that for $\alpha=1$ and Coulomb interactions the phase $(0,\pi/2)$ 
is absent and solely the phase $(\pi/4,\pi/4)$ exists for all $d\in[0,d_{c2}]$. 
As soon as $\alpha<1$, see Fig. \ref{fig:phida1}, the disordered phase $(0,0)$ 
takes place in the region of small $d\in [0,d_{c1}]$, with the standard
second-order transition to the symmetry-broken phase 
$(\varphi,\varphi)$ ($\varphi\ne 0$) at $d_{c1}>0$. 
We see from Fig. \ref{fig:phida1} that $d_{c1}\to 0$ as $\alpha\to 1$ and 
that $\varphi$ goes to $\pi/4$ in an asymptotic way. 

The above results can be reproduced analytically by using the representation
(\ref{e0}) of the energy $E_0$.
For small $d$, this energy can be expanded around $d\to 0$ as follows
\begin{eqnarray} 
E_0(\varphi,d) & = & \frac{\sqrt{\pi\alpha}}{2d} + 2 \Sigma_1(\alpha)
+\left[ \frac{8}{3\alpha}\pi^{3/2}+f_2(\varphi) \right] d^2 \nonumber\\
& &  +\left[ -\frac{16}{5\alpha^2}\pi^{5/2}+f_4(\varphi) \right] d^4
+ {\cal O}(d^6) . \label{e0exd}
\end{eqnarray}
The explicit formulas for the functions $f_2$ and $f_4$ are given in 
Eqs. (\ref{f2}) and (\ref{f4}), respectively.
Note the leading $1/d$ term, i.e., the dimer energy $1/(2d)$ renormalized 
by an infinite lattice summations of Coulomb energies. 
To minimize the energy (\ref{e0exd}) with respect to $\varphi$, we can 
forget about the $\varphi$-independent terms and it is all about 
$f_2(\varphi)d^2+f_4(\varphi)d^4$.
For $\alpha<1$, the $f_2$-term is relevant. 
It is expressible as
\begin{equation} \label{f2fit}
f_2(\varphi) = f_2\left(\textstyle{\frac{\pi}{4}}\right)
+\left[f_2(0)-f_2\left(\textstyle{\frac{\pi}{4}}\right) \right]
\cos{(2\varphi)} ,
\end{equation}
where the prefactor $f_2(0)-f_2(\pi/4)<0$ for any $\alpha<1$ and thus
the disordered phase $(0,0)$ minimizes the energy.

For $\alpha=1$ the prefactor $f_2(0)-f_2(\pi/4)$ vanishes and 
$f_2(\varphi)$ becomes constant, see \ref{C}.
Consequently, the function $f_4(\varphi)$ becomes relevant in 
the energy minimization.
According to \ref{C}, it can be expressed as
\begin{equation} \label{f4exact}
f_4(\varphi) = f_4\left(\textstyle{\frac{\pi}{8}}\right)
+\left[f_4(0)-f_4\left(\textstyle{\frac{\pi}{8}}\right)\right] 
\cos{(4\varphi)} , \qquad \alpha=1 .
\end{equation}
Since $f_4(0)-f_4(\pi/8) \approx 8.1946447789840126555$ 
is positive, the energy minimization requires that $4\varphi=\pi$ and 
we get the phase $(\pi/4,\pi/4)$.

\subsection{The second-order phase transition between the homogeneous phases}
\begin{figure}[]
\begin{center}
\includegraphics[clip,width=0.6\textwidth]{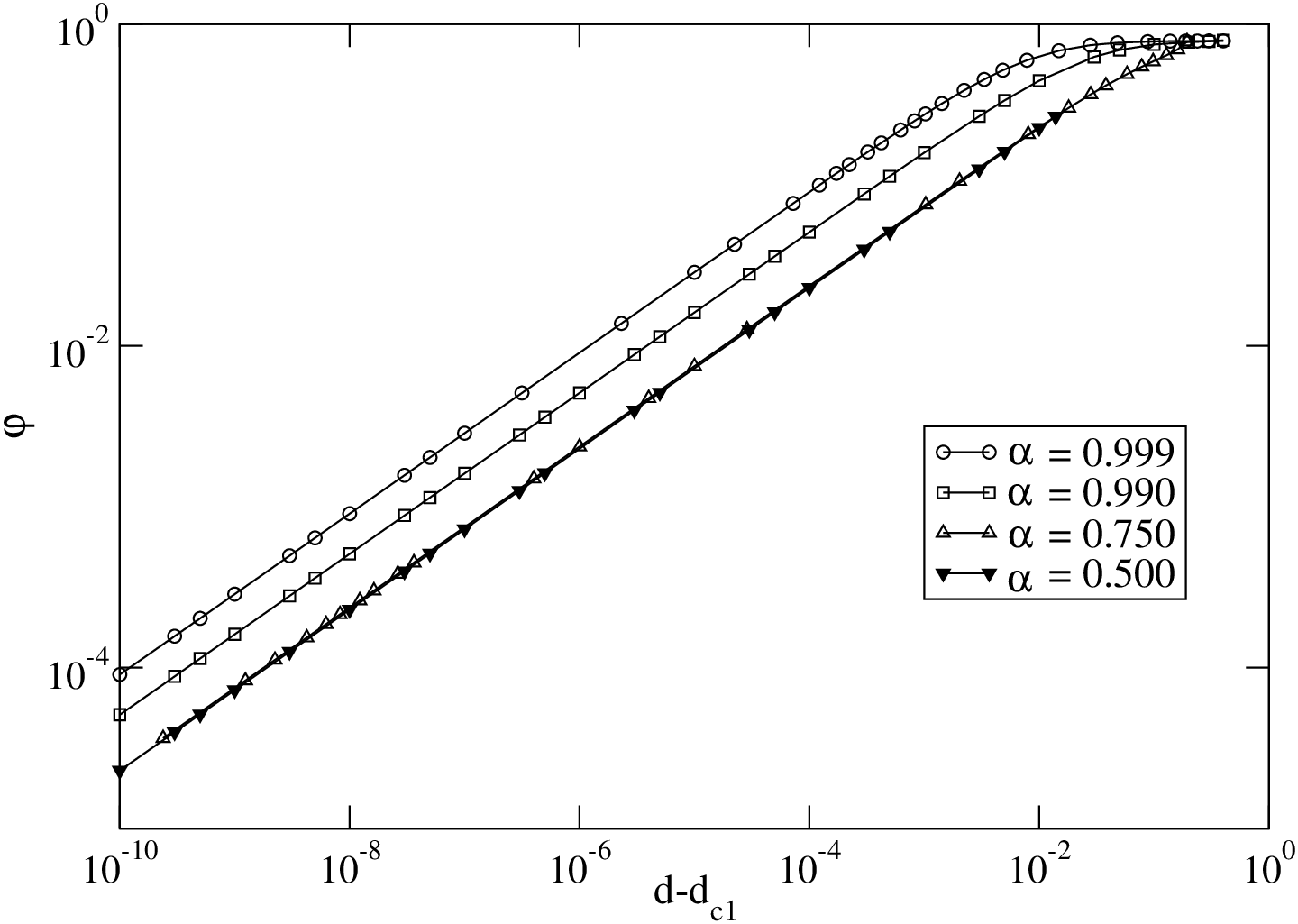}
\caption{The symmetry-broken angle $\varphi$ as a function of $d-d_{c1}$ 
for several values of $\alpha$, in the logarithmic scale.}
\label{fig:phid-dc}
\end{center}
\end{figure}

The critical point of the Landau expansion (\ref{e0ex}) is determined by
the standard condition \cite{Chaikin} 
\begin{eqnarray} \label{g20}
g_2(d_{c1}) = 0.
\end{eqnarray}
When approaching the critical point $d\to d_{c1}^+$, the order parameter 
$\varphi$ exhibits the singular behaviour of type
$\varphi\propto (d-d_{c1})^\beta$; in the mean-field approximation,
the exponent $\beta=1/2$. 
For various values of $\alpha$, we plot in Fig. \ref{fig:phid-dc}
numerical values of the order parameter $\varphi$ as the function
of the deviation from the critical point $d-d_{c1}$.   
In logarithmic scale, these dependences become linear for small values of 
$d-d_{c1}$ with all lines being parallel and thus having a common slope
$\beta$.
Numerical fits give $0.499999 \le\beta\le 0.500005$, confirming 
the mean-field value of the exponent $\beta$.

The mean-field behaviour can be derived also analytically.
The derivation is based on the Taylor expansions of the coefficients
$g_2$ and $g_4$ in (\ref{e0ex}) for small $d-d_{c1}$:
\begin{eqnarray} 
g_2(\alpha,d) = g_{21}(\alpha) (d-d_{c1}) + {\cal{O}}\left[(d-d_{c1})^2\right]
\nonumber\\
g_4(\alpha,d) = g_{40}(\alpha) + {\cal{O}}(d-d_{c1}). \label{g24taylor}
\end{eqnarray}
It was checked numerically that for any $\alpha$ it holds that
$g_{21}(\alpha)<0$ and $g_{40}(\alpha)>0$. 
The minimization of the energy (\ref{e0ex}) implies
\begin{equation} \label{de0dphi}
\frac{\partial E_0(\varphi,d)}{\partial\varphi}
= 2 g_2(\alpha,d) \varphi + 4 g_4(\alpha,d) \varphi^3 = 0 .
\end{equation}
The trivial solution $\varphi=0$ is dominant for $d\le d_{c1}$. 
For $d>d_{c1}$, we have the nontrivial
\begin{equation} \label{phid}
\varphi = \left( -\frac{g_2(\alpha,d)}{2g_4(\alpha,d)}\right)^{1/2}
\approx\left( -\frac{g_{21}(\alpha)}{2g_{40}(\alpha)}\right)^{1/2}
\sqrt{d-d_{c1}} ,
\end{equation}
with the mean-field exponent $\beta=1/2$.

The formula (\ref{g20}) yields the line $d_{c1}(\alpha)$, 
separating the phases $(0,0)$ and $(\varphi,\varphi)$ with $\varphi\ne 0$, 
see the phase diagram in Fig. \ref{fig:phasedia}.
This critical line is restricted by the condition $d<\alpha/2$ and 
the border value $d_{c1}=\alpha/2$ appears at $\alpha\approx 0.477648$.
The numerical fit indicates that $d_{c1}\to 0$ as $\alpha\to 1$  
according to $d_{c1}\propto \sqrt{1-\alpha}$.

\section{Second-order transition between the homogeneous and inhomogeneous 
phases} \label{Sec5}
As was explained in section \ref{Sec2}, there exists a special value of 
$\alpha^*$, separating two types of phase transitions between the homogeneous
($\varphi\ne 0,\delta\varphi=0$) and inhomogeneous 
($\varphi\ne 0,\delta\varphi\ne0$) phases.
In a large interval of $\alpha\in [\alpha^*,1]$, one has a second-order 
phase transition at $d_{c2}$, see Fig. \ref{f1a}, and in this section 
we shall concentrate on this phenomenon.

We start with the energy expansion (\ref{enex}).
Now we have two variational parameters and both partial derivatives 
of the energy with respect to $\varphi$ and $\delta\varphi$ must vanish.
The condition $\partial E/\partial (\delta\varphi)=0$ yields
\begin{equation} \label{deddp}
2 h_2(\varphi,d) \delta\varphi + 4 h_4(\varphi,d)(\delta\varphi)^3
+ \cdots = 0 .
\end{equation}
Besides the trivial disordered solution $\delta\varphi=0$ we have also 
the symmetry-broken solution $\delta\varphi\ge 0$ given by 
\begin{equation} \label{deddp2}
h_2(\varphi,d) + 2 h_4(\varphi,d)(\delta\varphi)^2 = 0 .
\end{equation}
The condition $\partial E/\partial \varphi=0$ leads to
\begin{equation} \label{dedp}
\frac{\partial E_0(\varphi,d)}{\partial\varphi} +
\frac{\partial h_2(\varphi,d)}{\partial\varphi}(\delta\varphi)^2 = 0 ,
\end{equation}
where the derivatives are taken at the physical value of $\varphi$.
Eqs. (\ref{deddp2}) and (\ref{dedp}) must be fulfilled simultaneously, 
so they can differ from one another only by a multiplicative factor $c$, 
\begin{equation} \label{h24de}
h_2(\varphi,d) = c\frac{\partial E_0(\varphi,d)}{\partial\varphi} , \qquad
2 h_4(\varphi,d) = c\frac{\partial h_2(\varphi,d)}{\partial\varphi} .
\end{equation}
Eliminating $c$ yields the equality
\begin{equation} \label{h24dedh2}
h_2(\varphi,d) \frac{\partial h_2(\varphi,d)}{\partial\varphi} =
2 h_4(\varphi,d) \frac{\partial E_0(\varphi,d)}{\partial\varphi} 
\end{equation}
whose validity was checked also numerically. 

\begin{figure}[]
\begin{center}
\includegraphics[clip,width=0.6\textwidth]{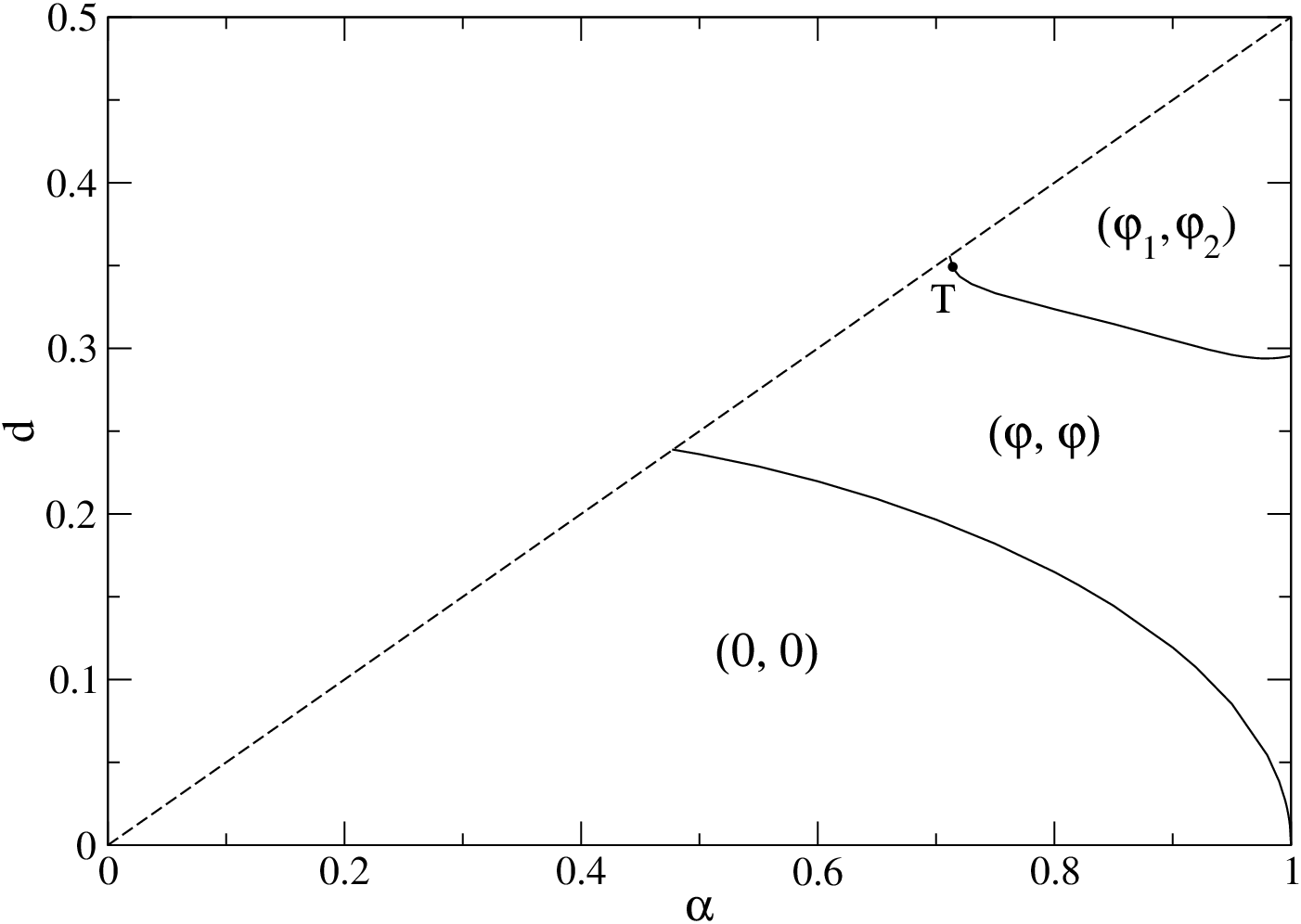}
\caption{The phase diagram of the homogeneous disordered $(0,0)$ and
symmetry-broken $(\varphi,\varphi)$ phases and the inhomogeneous 
$(\varphi_1,\varphi_2)$ phase. 
The dashed line marks the restriction $d\le \alpha/2$.
The short line localized upper-left from the tricritical point T 
corresponds to the first-order phase transitions.}
\label{fig:phasedia}
\end{center}
\end{figure}

To find the critical point, we insert the critical value $\delta\varphi=0$ 
into Eqs. (\ref{deddp2}) and (\ref{dedp}) to obtain
\begin{equation} \label{e0h2}
\frac{\partial E_0(\varphi,d_c)}{\partial\varphi}{\Bigg\vert}_{\varphi=\varphi_c} 
= 0 , \qquad h_2(\varphi_c;d_c) = 0 ;
\end{equation}
hereinafter, we use the simplified notation for $d_{c2}\equiv d_c$,
$\varphi(d_{c2})\equiv \varphi_c$, etc.
This set of two equations was used to calculate the pairs $(d_c,\varphi_c)$ 
for $\alpha\in [\alpha^*,1]$.
The critical line $d_{c2}(\alpha)$ between the homogeneous $(\varphi,\varphi)$
and inhomogeneous $(\varphi_1,\varphi_2)$ phases is represented 
in Fig. \ref{fig:phasedia}; it ends up at the tricritical point $T$ 
which corresponds to $\alpha=\alpha^*$ (specified later).

To find the dependence of the symmetry-broken $\delta\varphi$ on $d-d_c$
we note that there are two small variables in the vicinity of 
the critical point, namely $d-d_c$ and $\varphi-\varphi_c$.
We shall use the equality (\ref{h24dedh2}) and expand all functions
it contains in $d-d_c$ and $\varphi-\varphi_c$.
Respecting the critical condition (\ref{e0h2}), the expansion of 
$h_2(d,\varphi)$ around the critical point takes form
\begin{eqnarray} 
h_2(\varphi,d) & = & 
\frac{\partial h_2}{\partial \varphi}{\Bigg\vert}_c(\varphi-\varphi_c)
+ \frac{\partial h_2}{\partial d}{\Bigg\vert}_c (d-d_c)
+ \frac{1}{2}\Bigg[
\frac{\partial^2 h_2}{\partial d^2}{\Bigg \vert}_c(d-d_c)^2 \nonumber \\ & &
+\frac{\partial^2 h_2}{\partial \varphi^2}{\Bigg \vert}_c(\varphi-\varphi_c)^2
+2\frac{\partial^2 h_2}{\partial \varphi \partial d}{\Bigg \vert}_c(d-d_c)
(\varphi-\varphi_c)\Bigg] + \cdots, \label{h2exp}
\end{eqnarray}
where the symbol $\vert_c$ means at the critical point
($d\to d_c$ and $\varphi\to \varphi_c$). 
The function $h_4(\varphi,d)$ is expanded as
\begin{equation} \label{h4exp}
h_4(\varphi,d) = h_4(\varphi_c,d_c) + 
\frac{\partial h_4}{\partial\varphi}{\Bigg\vert}_c (\varphi-\varphi_c)+
\frac{\partial h_4}{\partial d}{\Bigg\vert}_c (d-d_c) .
\end{equation}
With respect to the critical condition (\ref{e0h2}), the expansion of 
$E_0(\varphi,d)$ reads
\begin{eqnarray} \label{e0exp}
E_0(\varphi,d) & = & E_0(\varphi_c,d_c)
+\frac{\partial E_0}{\partial d}{\Bigg\vert}_c (d-d_c)
+\frac{1}{2}\Bigg[\frac{\partial^2 E_0}{\partial d^2}{\Bigg\vert}_c (d-d_c)^2
\nonumber\\ & & 
+\frac{\partial^2 E_0}{\partial \varphi^2}{\Bigg\vert}_c (\varphi-\varphi_c)^2
+2 \frac{\partial^2 E_0}{\partial \varphi \partial d}{\Bigg\vert}_c 
(d-d_c)(\varphi-\varphi_c)\Bigg] \nonumber \\ & &
+\frac{1}{6}\Bigg[ \frac{\partial^3 E_0}{\partial \varphi^3}{\Bigg\vert}_c 
(\varphi-\varphi_c)^3 + 
3 \frac{\partial^3 E_0}{\partial \varphi^2 \partial d}{\Bigg\vert}_c 
(d-d_c)(\varphi-\varphi_c)^2+\cdots \Bigg] .
\end{eqnarray}

At the present stage, we can restrict ourselves to linear terms in 
Eqs. (\ref{h2exp}) and (\ref{e0exp}), but higher order terms will be 
important in the next section.
Inserting the expansions (\ref{h2exp}), (\ref{h4exp}) and (\ref{e0exp}) 
into the basic relation (\ref{h24dedh2}), we get
\begin{eqnarray} 
& & \Bigg[\frac{\partial h_2}{\partial d}{\Bigg\vert}_c (d-d_c)
+\frac{\partial h_2}{\partial \varphi}{\Bigg\vert}_c (\varphi-\varphi_c)\Bigg]
\frac{\partial h_2}{\partial \varphi}{\Bigg\vert}_c \nonumber \\
&& = 2 h_4(\varphi_c,d_c)\Bigg[ 
\frac{\partial^2 E_0}{\partial \varphi^2}{\Bigg\vert}_c (\varphi-\varphi_c)
+ \frac{\partial^2 E_0}{\partial \varphi \partial d}{\Bigg\vert}_c (d-d_c)
\Bigg] . \label{ddcppc}
\end{eqnarray}
Consequently,
\begin{equation} \label{ppcddc}
\varphi-\varphi_{c2} = a (d-d_{c2}), \qquad
a = \frac{\frac{1}{2h_4}\frac{\partial h_2}{\partial \varphi}
\frac{\partial h_2}{\partial d} -\frac{\partial^2 E_0}{
\partial \varphi \partial d}}{\frac{\partial^2 E_0}{\partial \varphi^2} -
\frac{1}{2h_4}\big(\frac{\partial h_2}{
\partial \varphi}\big)^2}{\Bigg \vert}_{c2},
\end{equation}
where $a$ has a non-zero denominator for $\alpha>\alpha^*$.
We see that the ``irrelevant'' field component $\varphi$, whose symmetry has 
already been broken starting from the previous critical point $d_{c1}(\alpha)$,
is an analytic function of $d-d_{c2}$.

To obtain the order parameter $\delta\varphi$, we apply the expansions
(\ref{h2exp}) and (\ref{ppcddc}) in Eq. (\ref{deddp2}), with the result
\begin{equation} \label{dp2}
(\delta\varphi)^2 = b (d-d_{c2}), \qquad
b = - \frac{1}{2h_4}\left[\frac{\partial h_2}{\partial d}
+ a \frac{\partial h_2}{\partial \varphi}\right]{\Bigg \vert}_{c2} ,
\end{equation}
where $b>0$ for $\alpha>\alpha^*$.
Thus the ``relevant'' field component 
$\delta\varphi = \sqrt{b} \sqrt{d-d_{c2}}$ exhibits the mean-field
critical behaviour with the exponent $\beta=1/2$.

\section{Tricritical point} \label{Sec6}
Within the standard Landau theory for the one-component field $\psi$ 
\cite{Chaikin}, the free energy $f(\psi)=f(-\psi)$ is written as
\begin{equation} \label{landau}
f(\phi) = g_2 \psi^2 + g_4 \psi^4 + g_6\psi^6 + \cdots ,
\end{equation}
where $g_2\sim g_{21} (T-T_t)$ ($g_{21}<0$ and $T_t$ is the transition 
temperature) and the next coefficients $g_{2n}$ depend on model's parameters.
If $g_4>0$, one gets a second-order transition at $g_2=0$.
For $g_4<0$, the first-order transition with a discontinuity in $\psi$ appears, 
for more details see Ref. \cite{Chaikin}.
In the $g_2-g_4$ plane, the lines of first-order and second-order
phase transitions merge at the tricritical point defined by $g_4=0$.
Provided that $g_6^*>0$ at the tricritical point $T^*$, the consequent 
minimization of the free energy (\ref{landau}) yields the non-trivial solution
\begin{eqnarray} \label{betatric}
\psi = \left( - \frac{g_2}{3g_6} \right)^{1/4}
\approx \left( - \frac{g_{21}}{3g_6^*} \right)^{1/4} (T-T^*)^{1/4} .
\end{eqnarray}
We see that the ordinary mean-field critical exponent $\beta=1/2$ jumps to 
$\beta=1/4$ at the tricritical point.

\bigskip
\begin{figure}[]
\begin{center}
\includegraphics[clip,width=0.6\textwidth]{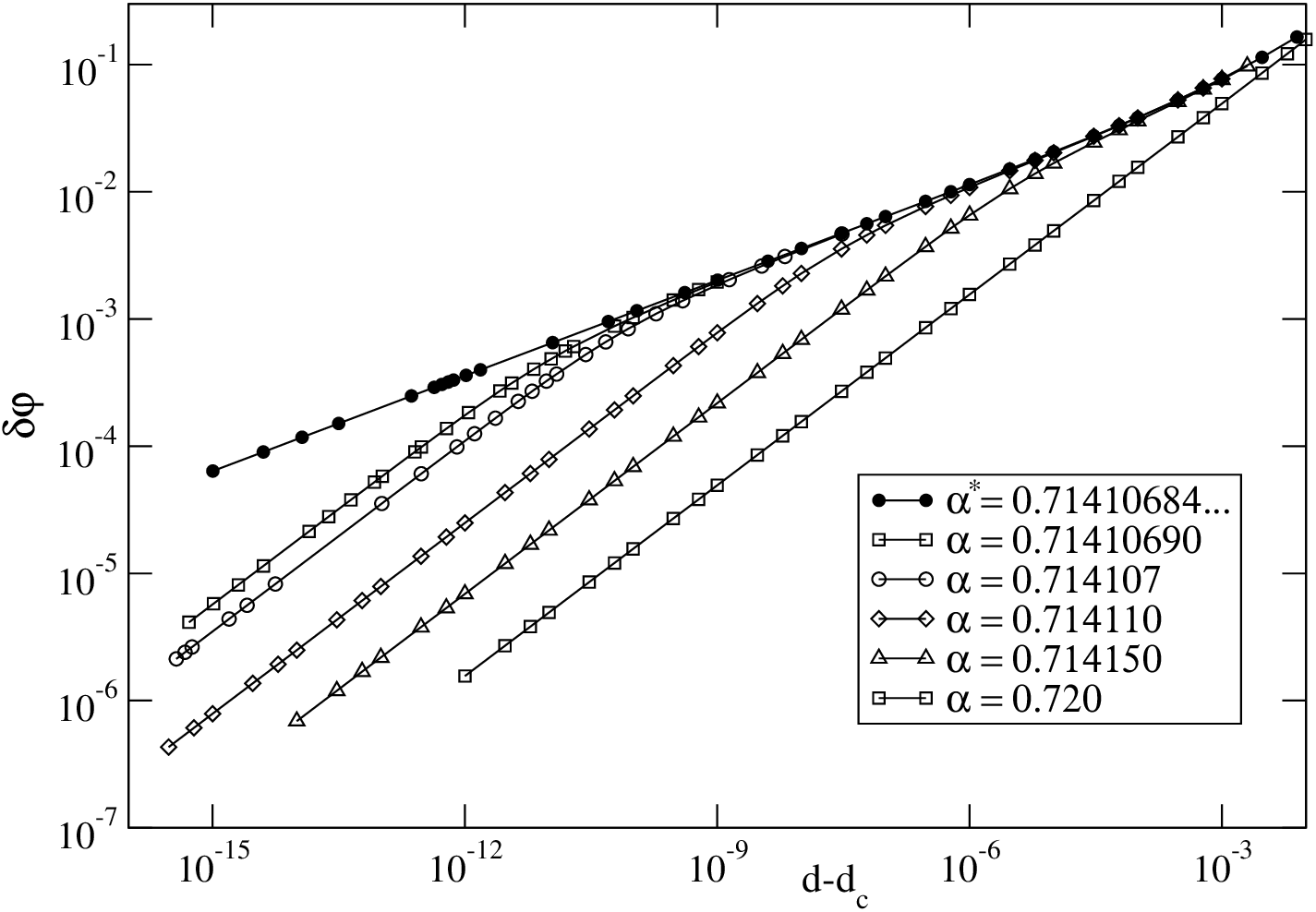}
\caption{Numerical results for the order parameter $\delta\varphi$ 
as the function of $d-d_c$ (in logarithmic scale), for several values of 
the aspect ration $\alpha\ge \alpha^*$.}
\label{fig:dphid-dc}
\end{center}
\end{figure}

Numerical calculations indicate that the prefactor $b$ in (\ref{dp2}) 
diverges when $\alpha\to(\alpha^*)^+$. 
The divergence of the prefactor is a typical signal of a change of 
the critical behaviour.
The coefficient $b$ from (\ref{dp2}) diverges when $a$ from (\ref{ppcddc}) 
diverges as well and this happens when the denominator in the expression for
$a$ becomes zero, i.e.,
\begin{equation} \label{den0}
\frac{\partial^2 E_0}{\partial \varphi^2}{\Bigg\vert}_c
- \frac{1}{2h_4(\varphi_c,d_c)}
\left(\frac{\partial h_2}{\partial \varphi}{\Bigg\vert}_c\right)^2
= 0 , \qquad \alpha=\alpha^* .
\end{equation}
Using this closed-form relation we were able to calculate the tricritical 
$\alpha^*$ with high precision, namely 
\begin{equation}
\alpha^*=0.71410684000071\ldots .
\end{equation}
The corresponding tricritical values of other parameters are 
\begin{eqnarray}
d^* & \equiv & d_{c2}(\alpha^*) = 0.3492349647792684\ldots, \nonumber \\
\varphi^* & \equiv & \varphi(\alpha^*,d^*) = 0.7134922360355926466\ldots .
\end{eqnarray}

The numerical plots of the order parameter $\delta\varphi$ versus $d-d_c$ 
are shown in logarithmic scale for various values of $\alpha\ge \alpha^*$
in Fig. \ref{fig:dphid-dc}. 
For $\alpha=\alpha^*$, we got a line with the expected slope 
$\beta^*\approx 0.250002$.
For $\alpha$-values slightly above $\alpha^*$ one can see in the plots
two regions with different slopes.
If $d-d_c$ is very small, the corresponding mean-field behaviour gives 
the slopes with $\beta$-values in between 0.499937 and 0.499954. 
If $d-d_c$ is large enough, since the $\alpha$-values do not differ much
from one another also the values of $\delta\varphi$ are close to each other 
and we have the tricritical $\beta^*\approx 0.250002$ slope. 
To obtain numerically the precise value of $\alpha^*$ one has to go to 
extremely small $d-d_c<10^{-12}$, hardly achievable in numerical simulations.

\begin{figure}[]
\begin{center}
\includegraphics[clip,width=0.6\textwidth]{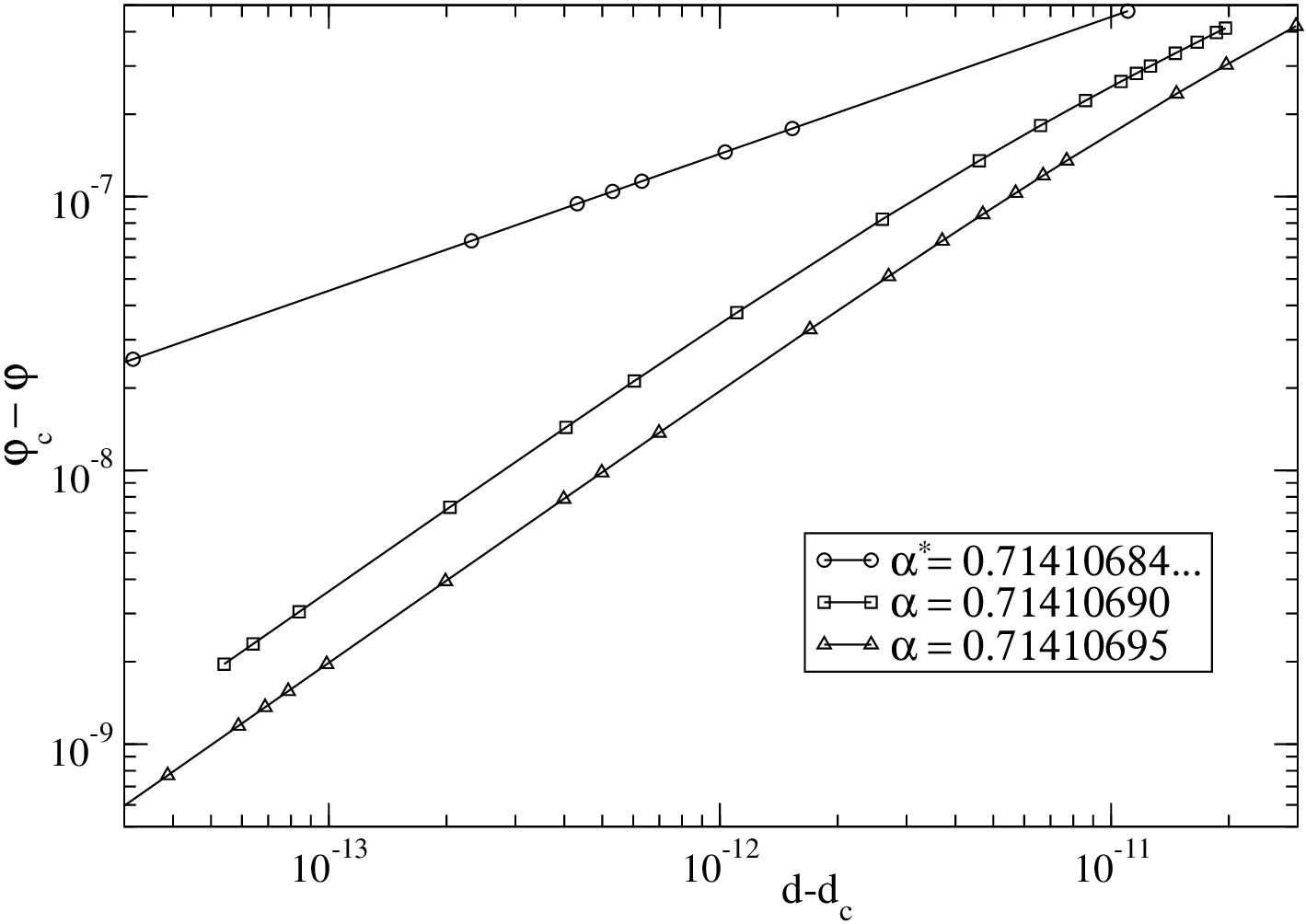}
\caption{Numerical results for the ``irrelevant'' quantity $\varphi_c-\varphi$ 
as the function of $d-d_c$ (in logarithmic scale), for several values of 
the aspect ration $\alpha\ge \alpha^*$.}
\label{fig:phiphic-dc}
\end{center}
\end{figure}

The simple condition $g_4=0$ from the standard one-component tricritical theory
does not apply to the present two-component theory, in particular our 
$h_4(\varphi^*,d^*)\ne 0$ at the tricritical point.
Our derivation of a relation similar to (\ref{betatric}) is connected with 
the existence of two quantities which vanish at the tricritical point: 
$\delta\varphi$ and $\varphi-\varphi_c$.
According to formula (\ref{ppcddc}), for $\alpha>\alpha^*$ 
the ``irrelevant'' quantity $\varphi-\varphi_c$, which has already broken 
its symmetry at $d_{c1}$, is an analytic function of $d-d_{c2}$. 
This is clearly seen in numerical data for the aspect ratio slightly above 
the tricritical one, namely $\alpha=0.7141069$ and $0.71410695$,
presented in Fig. \ref{fig:phiphic-dc}; the fitting ansatz 
$\varphi_c-\varphi\propto (d-d_c)^{\zeta}$ (the exponent is denoted as $\zeta$
because $\varphi_c-\varphi$ {\em is not} an order parameter) gives the values 
$\zeta = 0.996$ and $0.99995$ respectively.
But at the tricritical point the numerical fit implies 
$\zeta^*\approx 0.499998$, i.e., also $\varphi-\varphi_c$ surprisingly 
exhibits singular critical behaviour, presumably of mean-field type. 

To reproduce the above numerical findings also analytically, in full analogy 
with the previous section we insert the expansions (\ref{h2exp}), 
(\ref{h4exp}) and (\ref{e0exp}) into the relation (\ref{h24dedh2}),
keeping also certain relevant higher-order terms:
\begin{eqnarray} 
& & \left[\frac{\partial h_2}{\partial d}{\Bigg\vert}_c (d-d_c)
+ \frac{\partial h_2}{\partial\varphi}{\Bigg\vert}_c (\varphi-\varphi_c)
+\frac{\partial^2 h_2}{\partial \varphi^2}{\Bigg\vert}_c
(\varphi-\varphi_c)^2 \right] \nonumber \\
& & \quad \times 
\left[\frac{\partial h_2}{\partial \varphi}{\Bigg\vert}_c
+\frac{\partial^2 h_2}{\partial \varphi^2}{\Bigg\vert}_c (\varphi-\varphi_c)
\right] \nonumber \\
& & = 2\left[ h_4(d_c,\varphi_c)
+ \frac{\partial h_4}{\partial \varphi}{\Bigg\vert}_c (\varphi-\varphi_c)
\right] \nonumber \\ & & \times \left[ 
\frac{\partial^2 E_0}{\partial \varphi^2}{\Bigg\vert}_c (\varphi-\varphi_c)
+\frac{\partial^2 E_0}{\partial\varphi\partial d}{\Bigg\vert}_c (d-d_c)
+\frac{1}{2}\frac{\partial^3 E_0}{\partial \varphi^3}{\Bigg\vert}_c
(\varphi-\varphi_c)^2\right]. \label{ddcppc2}
\end{eqnarray}
At $\alpha=\alpha^*$, after expanding the brackets the two terms linear in 
$\varphi-\varphi_c$ cancel with one another due to the tricritical
constraint (\ref{den0}).
This cancellation plays a similar role as the condition $g_4 = 0$ in 
the standard one-component theory of tricriticality.
Thus the terms of order $d-d^*$ and $(\varphi-\varphi^*)^2$ become
the leading ones and we arrive at the result 
\begin{equation} \label{pddc}
\varphi-\varphi^* = a^* \sqrt{d-d^*} , \qquad
a^* = - \sqrt{\frac{\frac{1}{2}\frac{\partial h_2}{\partial d}
\frac{\partial h_2}{\partial\varphi} -
h_4\frac{\partial^2 E_0}{\partial \varphi \partial d}}
{\frac{1}{2}h_4\frac{\partial^3 E_0}{\partial \varphi^3}
+ \frac{\partial h_4}{\partial \varphi}\frac{\partial^2 E_0}{\partial \varphi^2}
- \frac{\partial^2 h_2}{\partial \varphi^2}
\frac{\partial h_2}{\partial \varphi}}}\ {\Bigg \vert}^* ,
\end{equation}
where the symbol $\vert^*$ means at $(d^*,\varphi^*)$. 
The minus sign in the definition of $a^*$ is fixed by numerical results 
in Fig. \ref{fig:phiphic-dc}, the numerical value of 
$a^*\approx -0.14073465315$.
We conclude that the ``irrelevant'' quantity $\varphi-\varphi^*$ indeed
exhibits a mean-field singularity at the tricritical point.

We calculate $(\delta\varphi)^2$ again by applying (\ref{h2exp})
and (\ref{pddc}) in (\ref{deddp2}):
\begin{eqnarray} 
(\delta\varphi)^2 = -\frac{h_2(\varphi,d)}{2h_4(\varphi,d)} 
=-\frac{1}{2h_4}\frac{\partial h_2}{\partial\varphi}{\Bigg\vert}^*
(\varphi-\varphi^*) = b^* \sqrt{d-d^*}, \nonumber\\ 
b^* = - a^* \frac{1}{2h_4}\frac{\partial h_2}{\partial\varphi}{\Bigg \vert}^*
\approx 0.1258476191835. \label{dpddc2}
\end{eqnarray}
Finally we get $\delta\varphi=\sqrt{b^*}(d-d^*)^{1/4}$, so $\beta^*=1/4$ 
as was anticipated.

\section{First-order phase transitions for $\alpha<\alpha^*$} \label{Sec7}
Respecting the constraint (\ref{constraint}), there is a short interval 
of $\alpha\in [0.711535,\alpha^*]$ when at some transition value $d_t$ 
the quantity $\delta\varphi$ exhibits a discontinuous change from zero 
to a non-zero value, which is typical for first-order phase transitions.
We have not at our disposal an analytic theory for such transitions.
The values $\varphi_1$ and $\varphi_2$ are calculated by numerical 
minimization of the energy (\ref{en}).
In this way we got the first-order transition line $d_t(\alpha)$ 
for $\alpha$ from the above mentioned interval, see the short line 
localized upper-left from the tricritical point T in Fig. \ref{fig:phasedia}.
Close to the tricritical point it holds numerically that
\begin{equation} \label{swap}
\alpha^* - \alpha \approx 0.21134 (d_t - d^*) , \qquad
\alpha^* - \alpha, d_t - d^* \ll 1 . 
\end{equation}

\begin{figure}[]
\begin{center}
\includegraphics[clip,width=0.6\textwidth]{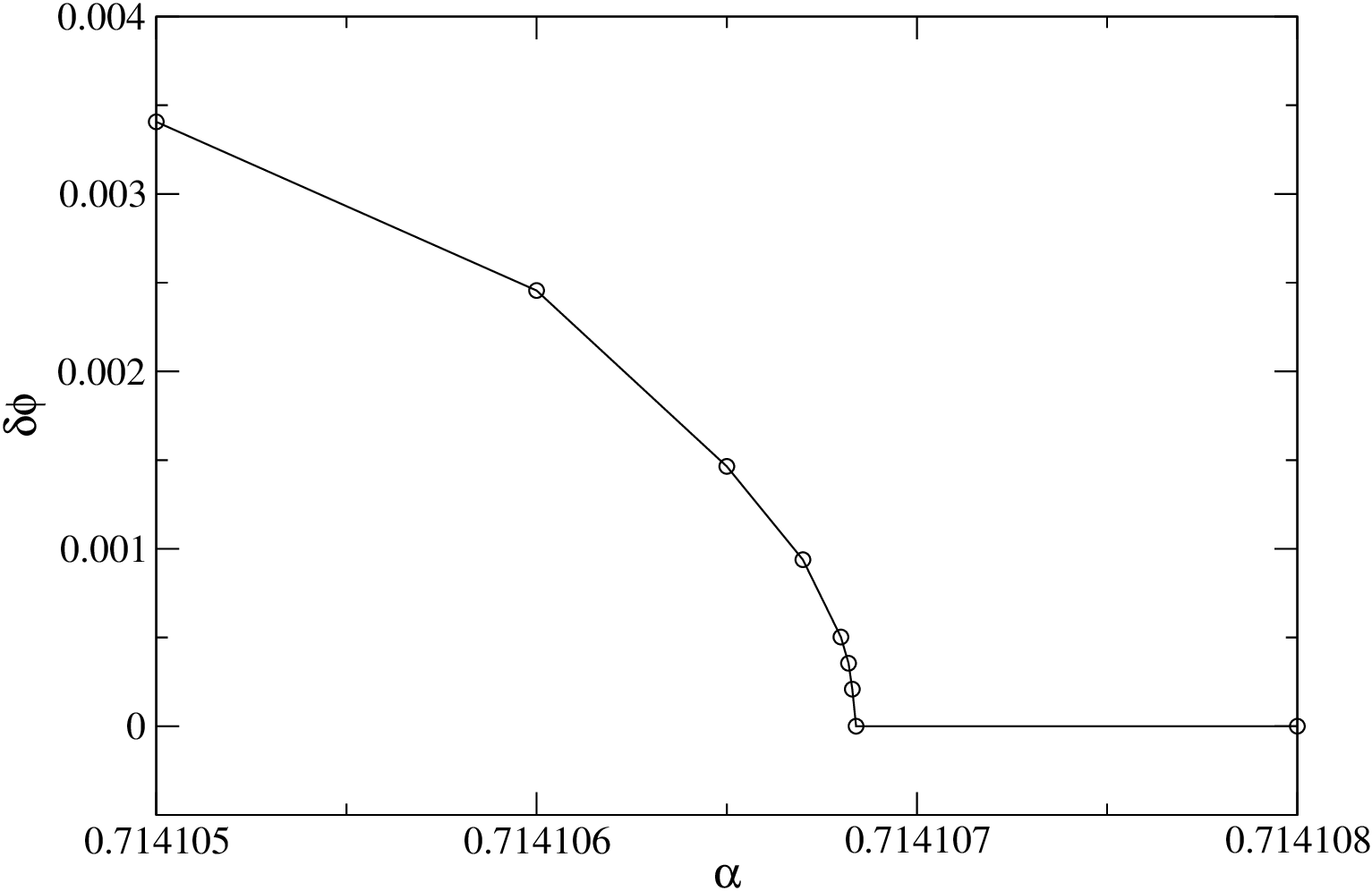}
\caption{The discontinuity of the difference between sublattice angles 
$\delta\phi$ at the first-order transition point $d_t$ as a function of 
the aspect ratio $\alpha$.
$\delta\phi=0$ for $\alpha\ge \alpha^*$}
\label{fig:deltaphi-a}
\end{center}
\end{figure}

Both angles $\varphi_1$ and $\varphi_2$, or equivalently $\varphi$
and $\delta\varphi$, exhibit discontinuities at $d_t(\alpha)$.
To describe these discontinuities, we introduce the quantities
\begin{equation}
\phi(\alpha) = \varphi(\alpha^-) - \varphi(\alpha^+)
\end{equation}
and
\begin{equation}
\delta\phi(\alpha) = \delta\varphi(\alpha^-) ,
\end{equation}
where we have taken into account that $\delta\varphi(\alpha^+)=0$.
We calculated $\delta\phi$ for several values of $\alpha$, see 
Fig. \ref{fig:deltaphi-a}.
The empirical fit gives $\delta\phi\propto (\alpha^* - \alpha)^{\tau}$ 
with $\tau\approx 0.4999$.
On the other hand, the function $\phi(\alpha)$ is well fitted by 
the linear dependence on $(\alpha^* - \alpha)$.

It is possible to analyze the dependence of $\phi$ and $\delta\phi$
on $d_t-d^*$, instead of $\alpha^*-\alpha$.
Due to the linear relation (\ref{swap}) the exponents remain unchanged.

\section{Concluding remarks} \label{Sec8}
We have studied the system of rotating dimers which consist of two 
equivalent Coulomb charges at distance $2d$. 
Dimer centers are localized on sites of a rectangular lattice with 
the aspect ratio $\alpha\le 1$.
The ground-state energy of such system is expressible in terms of two
components: $\varphi$, controlling the second-order phase transition
between the disordered and symmetry-broken homogeneous phases, and 
$\delta\varphi$, controlling the first-order or second-order phase transition
between the homogeneous and spatially inhomogeneous phases.
Our method of lattice summation ensures an extreme precision
of numerical results.
We were able to perform analytically Landau-type expansions of
the energy per particle in the corresponding order parameter, 
where the coefficient are expressed as infinite series of Misra functions.
This enabled us to determine the critical and tricritical points as
solutions of closed-form relations; the only approximation is the order
of the truncation of the Misra-function series.
The expansion of the Misra functions close to the (tri)critical point 
permits one to extract analytically the singular expansion of 
the order parameter. 
The main result is the observation of the tricritical point, separating 
the line of second-order phase transitions from the line of first-order ones
between the homogeneous and inhomogeneous phases.
The precise location of the tricritical point is a serious problem in
numerical methods due to the extreme need of precision.
We have shown both numerically and analytically that for second-order
phase transitions the critical exponent $\beta$ has its mean-field value $1/2$,
except for the tricritical point where it jumps to $\beta^*=1/4$.
Such phenomenon is known from literature about the Landau theory of 
one-component fields, but in our two-component version the scenario is 
more complicated: the already symmetry-broken component $\varphi$ 
(which therefore does not undertake the symmetry breaking at the tricritical 
point) also interferes and exhibits surprisingly the mean-field singular 
behaviour. 

Our phase diagram for the Coulomb interaction in Fig. \ref{fig:phasedia} 
is similar to the one in Ref. \cite{Trizac10} for the Yukawa interaction,
with two important differences. 
Firstly, no tricritical point was reported in \cite{Trizac10}.
The authors performed calculations just for several values of $\kappa$ and 
it is likely that the tricriticality is not intrinsic only to 
the Coulomb $\kappa\to 0$ limit, but it survives also for non-zero 
(at least small) $\kappa$-values.
The second difference concerns the square-lattice limit $\alpha\to 1$.
The boundary between phases $(\varphi,\varphi)$ and $(\varphi_1,\varphi_2)$ 
is localized at a non-zero value of $d_{c2}\approx 0.29542221$ in 
the Coulomb case, whereas $d_{c2}$ tends to zero in the Yukawa case.
In the Yukawa case, there exists a small region of the $(0,\pi/2)$ phase 
which we did not observe.

\begin{figure}[]
\begin{center}
\includegraphics[clip,width=0.6\textwidth]{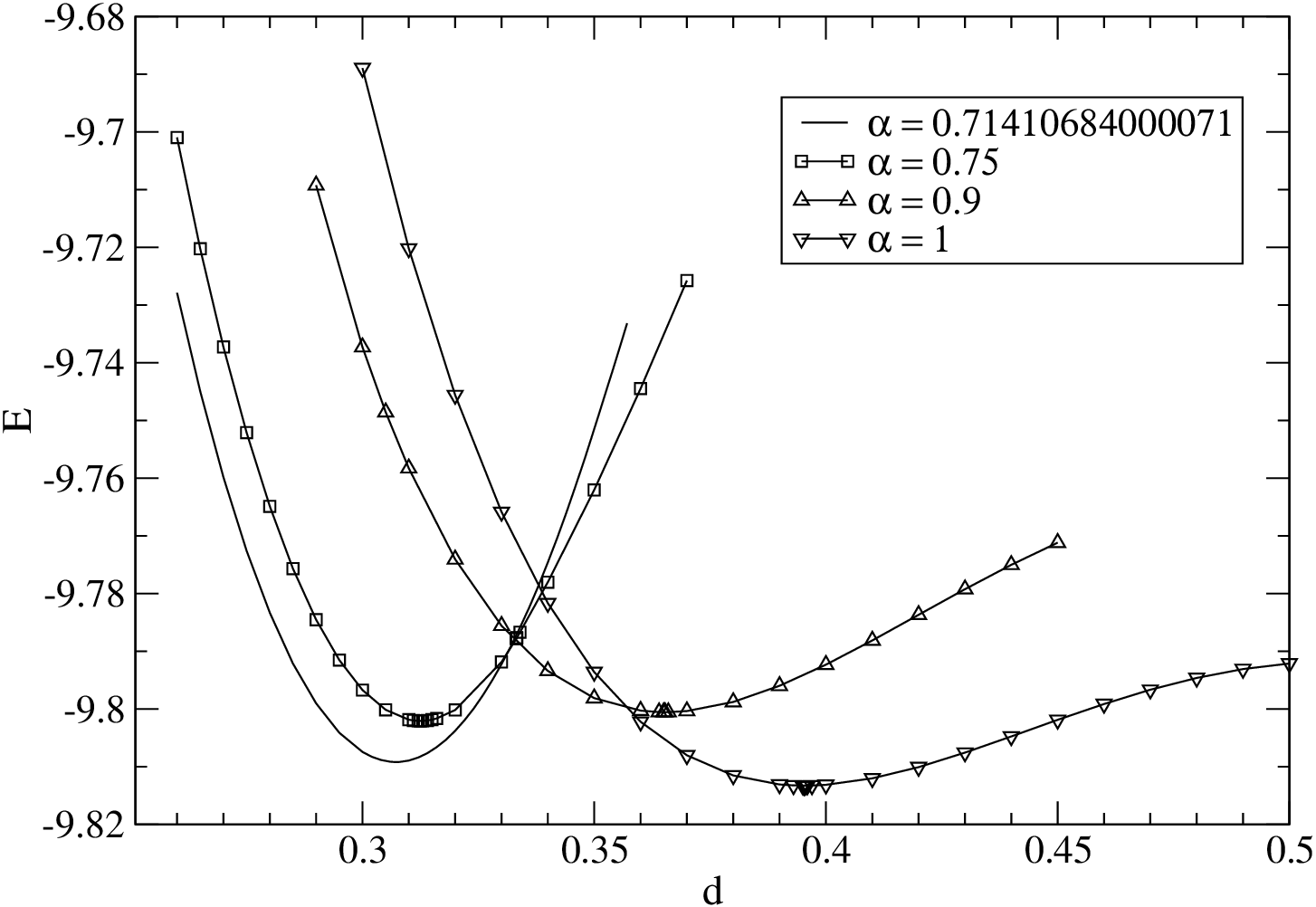}
\caption{Energy per particle as the function of $d$ for several values of 
$\alpha\ge\alpha^*$.}
\label{fig:energy-d}
\end{center}
\end{figure}

Finally, in Fig. \ref{fig:energy-d} we present for several values of 
the aspect ratio $\alpha\ge\alpha^*$ the energy per particle as a function 
of $d$.
For each $\alpha$, the function $E(d)$ has just one minimum at 
$d_{\min}(\alpha)$.
The smallest minimum occurs for $\alpha=1$ and it corresponds to
$d_{\min}(1)>d_{c2}(1)$, i.e., to the inhomogeneous phase with 
$\varphi>0$ and $\delta\varphi>0$.
Decreasing $\alpha$ leads to an increase in the energy minimum, up to 
a certain value of $\alpha$ when the energy minimum starts to decrease.
At $\alpha=\alpha^*$, the minimum occurs at $d_{\min}(\alpha^*)$ from 
the interval $d_{c1}(\alpha^*)<d_{\min}(\alpha^*)<d_{c2}(\alpha^*)$, i.e., 
with the homogeneous phase characterized by $\varphi>0$. 
This phenomenon causes the non-monotonic behaviour of the minimum energy 
as $\alpha$ decreases from 1 to $\alpha^*$. 

For future investigations it would be interesting to study whether 
the existence of the tricritical point is related to Coulomb interactions
or it can be found also for Yukawa interactions, at least in the
region of small $\kappa$.
Other systems like Coulomb dimers or dipoles on the triangular lattice 
\cite{Trizac10} are also of potential interest.

\ack
We thank Emmanuel Trizac for valuable discussions about the phenomenon
of tricriticality.
The support received from the grant VEGA No. 2/0015/2015 is acknowledged. 

\appendix

\section{Misra functions} \label{A}
In this paper, we work with Misra functions (\ref{Misra}) of half-integer 
index $\nu$.
They are expressible in terms of the complementary error function 
\cite{Gradshteyn}
\begin{equation} \label{erfc}
{\rm erfc}(z)=\frac{2}{\sqrt{\pi}}\int_z^\infty {\rm d}t\, \exp(-t^2)
\end{equation}
as follows \cite{Travenec15}
\begin{eqnarray}
z_{3/2}(y) & = & \sqrt{\frac{\pi}{y}}\ {\rm erfc}\left(\sqrt{\pi y}\right),
\nonumber \\
z_{5/2}(y) & = & \frac{\sqrt{\pi}}{2 y^{3/2}}\left[ 2 {\rm e}^{-\pi y}\sqrt{y} 
+{\rm erfc}{\left(\sqrt{\pi y}\right)} \right] , \nonumber \\
z_{7/2}(y) & = & \frac{\sqrt{\pi}}{4 y^{5/2}}
\left[ 2 {\rm e}^{-\pi y}\sqrt{y}\left(3+2\pi y\right) 
+ 3 {\rm erfc}{\left(\sqrt{\pi y}\right)} \right] , \nonumber \\ 
z_{9/2}(y) & = & \frac{\sqrt{\pi}}{8 y^{7/2}}
\left\{ 2{\rm e}^{-\pi y}\sqrt{y} (15 + 10 \pi y  + 4 \pi^2 y^2) 
+ 15 {\rm erfc}{\left(\sqrt{\pi y}\right)}\right\}, \nonumber \\
z_{11/2}(y) & = & \frac{\sqrt{\pi}}{16 y^{9/2}}
\big\{ 2{\rm e}^{-\pi y}\sqrt{y} (105 + 70 \pi y + 28 \pi^2 y^2 + 8 \pi^3 y^3) 
\nonumber \\ & & 
+ 105 {\rm erfc}{\left(\sqrt{\pi y}\right)} \big\} . \label{znuy}
\end{eqnarray}

The expansion of $z_{\nu}(y+\delta y)$ for small $\delta y$ reads as
\cite{Travenec15,Travenec16}
\begin{eqnarray}
z_{\nu}(y+\delta y) & = & z_\nu(y) -\delta y\ z_{\nu+1}(y) 
+ \frac{(\delta y)^2}{2}\ z_{\nu+2}(y)\nonumber \\
& & - \frac{(\delta y)^3}{6}\ z_{\nu+3}(y)
+\frac{(\delta y)^4}{24}\ z_{\nu+4}(y)-\cdots . \label{zexp}
\end{eqnarray}

We also need the expansion of $z_{3/2}(y)$ for small $y$:
\begin{equation} \label{z32ex}
z_{3/2}(y) = \frac{\sqrt{\pi}}{y} - 2\sqrt{\pi} + \frac{2}{3}\pi^{3/2}y
-\frac{5}{3}\pi^{3/2}y^2 + \cdots ,
\end{equation}
where we applied the well known expansion of 
the error function \cite{Gradshteyn}.

\section{The expansions of the energy} \label{B}
Using the expansion formula for the Misra functions (\ref{zexp}), 
the energy $E_0$ (\ref{e0}) can be expanded as (\ref{e0ex}), where
\begin{eqnarray} 
g_2(d) & = & 2 \sum_{j=1}^{\infty} \left[ 2 d j \pi\sin(4dj\pi) z_{3/2}(j^2\alpha) 
- \frac{8d^2 j^2\pi^2}{\alpha^2} 
z_{3/2}\left(\textstyle{\frac{j^2}{\alpha}}\right) \right] \nonumber \\
& & +4 \sum_{j,k=1}^\infty \left[ 2dj\pi\sin(4dj\pi) -
\frac{8d^2 k^2\pi^2\cos(4dj\pi)}{\alpha^2}\right]
z_{3/2}\left(\textstyle{j^2\alpha+\frac{k^2}{\alpha}}\right) \nonumber \\
& & + \frac{2d}{\alpha} \sum_{j,k=-\infty}^\infty j
z_{5/2}\left[\textstyle{\frac{(j+2d)^2}{\alpha}+k^2\alpha}\right] \nonumber \\
& & + 8 d^2 \sum_{j,k=-\infty}^\infty k^2
z_{7/2}\left[\textstyle{\frac{(j+2d)^2}{\alpha}+k^2\alpha}\right] \label{g2}
\end{eqnarray}
and
\begin{eqnarray} 
g_4(d) & = & 2 \sum_{j=1}^\infty \left[-2d^2j^2\pi^2\cos(4dj\pi)
- \frac{d j \pi}{6} \sin(4dj\pi) \right] z_{3/2}(j^2\alpha) \nonumber \\
& & + \frac{16}{3\alpha^4} \sum_{j=1}^\infty (d^2j^2\pi^2\alpha^2+4d^4j^4\pi^4)
z_{3/2}\left(\textstyle{\frac{j^2}{\alpha}}\right) \nonumber \\
& & + 4 \sum_{j,k=1}^\infty \Bigg\{\left[
\frac{8(d^2k^2\pi^2\alpha^2+4d^4k^4\pi^4)}{3\alpha^4}
-2d^2j^2\pi^2\right]\cos(4dj\pi) \nonumber\\ 
& & - \left[\frac{1}{6}dj\pi+\frac{16d^3jk^2\pi^3}{\alpha^2}\right]
\sin(4dj\pi) \Bigg\}
z_{3/2}\left(\textstyle{j^2\alpha+\frac{k^2}{\alpha}}\right) \nonumber\\ 
& & - \frac{d}{6\alpha} \sum_{j,k=-\infty}^\infty j 
z_{5/2}\left[\textstyle{\frac{(j+2d)^2}{\alpha}+k^2\alpha}\right]\nonumber\\
& & + \sum_{j,k=-\infty}^\infty \left(\frac{2d^2j^2}{\alpha^2}
-\frac{8d^2k^2}{3}\right)
z_{7/2}\left[\textstyle{\frac{(j+2d)^2}{\alpha}+k^2\alpha}\right] \nonumber \\
& & + \frac{16d^3}{\alpha} \sum_{j,k=-\infty}^\infty j k^2 
z_{9/2}\left[\textstyle{\frac{(j+2d)^2}{\alpha}+k^2\alpha}\right] \nonumber \\
& & + \frac{32d^4}{3} \sum_{j,k=-\infty}^\infty k^4 
z_{11/2}\left[\textstyle{\frac{(j+2d)^2}{\alpha}+k^2\alpha}\right]. \label{g4}
\end{eqnarray}

Similarly, using (\ref{zexp}) the most general energy (\ref{en}) 
can be expanded in $\delta\varphi$ according to (\ref{enex}), with
the coefficient
\begin{eqnarray} \label{h2}
h_2(d) & = & - 2 d^2 \pi^2 \sum_{j=1}^\infty (-1)^j j^2 \sin^2\varphi\,
z_{3/2}(j^2\alpha) - \frac{2 d^2\pi^2\cos^2{\varphi}}{\alpha^2} \sum_{j=1}^\infty 
j^2 z_{3/2}\left(\textstyle{\frac{j^2}{\alpha}}\right) \nonumber \\
& & - 4 d^2\pi^2 \sum_{j,k=1}^\infty \left( \frac{k^2\cos^2{\varphi}}{\alpha^2}
+ j^2 \sin^2\varphi \right) (-1)^j
z_{3/2}\left(\textstyle{j^2\alpha+\frac{k^2}{\alpha}}\right) \nonumber \\
& & - \frac{d^2}{2\alpha} \sum_{j,k=-\infty}^\infty 
z_{5/2}\left[\textstyle{\frac{(j+1/2)^2}{\alpha}+k^2\alpha} \right] \nonumber \\
& & + d^2 \sum_{j,k=-\infty}^\infty 
\frac{[2\alpha k \cos{\varphi}- (1+2j)\sin{\varphi}]^2}{4\alpha^2}
z_{7/2}\left[\textstyle{\frac{(j+1/2)^2}{\alpha}+k^2\alpha}\right] \nonumber \\
& & + d\pi\sum_{j=1}^\infty \Biggl[ j\cos{\varphi}
\sin{(j\pi+2dj\pi\cos{\varphi})} z_{3/2}(j^2\alpha) \nonumber \\
& & + \frac{j\sin{\varphi}}{\alpha}
\sin{\left(\textstyle{\frac{2 d j\pi\sin{\varphi}}{\alpha}}\right)}
z_{3/2}\left(\textstyle{\frac{j^2}{\alpha}}\right) \Biggr] \nonumber \\
& & + 2 d\pi \sum_{j,k=1}^\infty \Bigg[ j\cos{\varphi}
\cos{\left(\textstyle{\frac{2 d k\pi\sin{\varphi}}{\alpha}}\right)}
\sin{(j\pi+2dj\pi\cos{\varphi})} \nonumber \\ & & +\frac{k}{\alpha}
\cos{(j\pi+2dj\pi\cos{\varphi})} \sin{\varphi}
\sin{\left(\textstyle{\frac{2 d k\pi\sin{\varphi}}{\alpha}}\right)}
\Bigg] z_{3/2}\left(\textstyle{j^2\alpha+\frac{k^2}{\alpha}} \right) \nonumber \\
& & + \frac{d}{2\alpha} \sum_{j,k=-\infty}^\infty \left[
\cos{\varphi}\left(\frac{1}{2}+j+d\cos{\varphi}\right)
+ \sin{\varphi} \left(\alpha k+d\sin{\varphi}\right) \right] \nonumber\\ 
& & \times 
z_{5/2}\left[\textstyle{\frac{\left(1/2+j+d\cos{\varphi}\right)^2}{\alpha}+
\left(k+\frac{d\sin{\varphi}}{\alpha}\right)^2\alpha}\right] \nonumber\\
& & - 2 d^2\pi^2 \sum_{j=1}^\infty \Bigg[ j^2 \sin^2{\varphi}\,
z_{3/2}(j^2\alpha) + \frac{(-1)^j j^2 \cos^2{\varphi}}{\alpha^2}
z_{3/2}\left(\textstyle{\frac{j^2}{\alpha}}\right)\Bigg] \nonumber \\
& & - 4 d^2\pi^2 \sum_{j,k=1}^\infty \left[\frac{k^2\cos^2{\varphi}}{\alpha^2}
+ j^2 \sin^2{\varphi}\right] (-1)^k 
z_{3/2}\left(\textstyle{j^2\alpha+\frac{k^2}{\alpha}}\right) \nonumber \\
& & - \frac{d^2}{2\alpha} \sum_{j,k=-\infty}^\infty 
z_{5/2}\left[\textstyle{\frac{j^2}{\alpha}+\left(k+\frac{1}{2}\right)^2\alpha}
\right] \nonumber \\
& & + \frac{d^2}{4} \sum_{j,k=-\infty}^\infty 
\left[(1+2k)\cos{\varphi}-\frac{2j\sin{\varphi}}{\alpha}\right]^2
z_{7/2}\left[\textstyle{\frac{j^2}{\alpha}+\left(k+\frac{1}{2}\right)^2\alpha}
\right] \nonumber \\ & & + d\pi \sum_{j=1}^\infty \Bigg[
j \cos{\varphi} \sin{\left(2dj\pi\cos{\varphi}\right)} z_{3/2}(j^2\alpha)
\nonumber\\ & & + \frac{j\sin{\varphi}}{\alpha}
\sin{\left(\textstyle{j\pi+\frac{2dj\pi\sin{\varphi}}{\alpha}}\right)}
z_{3/2}\left(\textstyle{\frac{j^2}{\alpha}}\right) \Bigg] \nonumber \\
& & + 2 d\pi \sum_{j,k=1}^\infty \Bigg[ j\cos{\varphi}
\cos{\left(\textstyle{k\pi+\frac{2dk\pi\sin{\varphi}}{\alpha}}\right)}
\sin{\left(2dj\pi\cos{\varphi}\right)} \nonumber\\
& & +\frac{k}{\alpha}\cos{\left(2dj\pi\cos{\varphi}\right)} \sin{\varphi}
\sin{\left(\textstyle{k\pi+\frac{2dk\pi\sin{\varphi}}{\alpha}}\right)}
\Bigg] z_{3/2}\left(\textstyle{j^2\alpha+\frac{k^2}{\alpha}}\right) \nonumber\\
& & + \frac{d}{2} \sum_{j,k=-\infty}^\infty \left[
\frac{\cos{\varphi}\ (j+d\cos{\varphi})}{\alpha} + \sin{\varphi}
\left( \frac{1}{2}+k+\frac{d\sin{\varphi}}{\alpha}\right)\right] \nonumber \\
& & \times z_{5/2}\left[\textstyle{\frac{(j+d\cos{\varphi})^2}{\alpha}+\alpha
\left(\frac{1}{2}+k+\frac{d\sin{\varphi}}{\alpha}\right)^2}\right] \nonumber \\
& & + \sum_{j=1}^\infty \Bigg\{ \big[ -2 d^2 j^2 \pi^2
\cos{\left(2dj\pi\cos{\varphi}\right)}\sin^2{\varphi} \nonumber \\
& & + dj\pi\cos{\varphi}\sin{\left(2dj\pi\cos{\varphi}\right)} \big] 
z_{3/2}(j^2\alpha) \nonumber \\ & & + \frac{1}{\alpha^2} \Bigg[ - 2 d^2 j^2 \pi^2
\cos^2{\varphi}
\cos{\left(\textstyle{\frac{2dj\pi\sin{\varphi}}{\alpha}}\right)} \nonumber\\
& & + \alpha dj\pi \sin{\varphi}
\sin{\left(\textstyle{\frac{2dj\pi\sin{\varphi}}{\alpha}}\right)} 
\Bigg] z_{3/2}\left(\textstyle{\frac{j^2}{\alpha}}\right) \Bigg\} \nonumber\\
& & + 2 \sum_{j,k=1}^\infty \Bigg[ -\frac{2d^2k^2\pi^2\cos^2{\varphi}}{\alpha^2}
\cos{\left(2dj\pi\cos{\varphi}\right)}
\cos{\left(\textstyle{\frac{2dk\pi\sin{\varphi}}{\alpha}}\right)} \nonumber\\
& & - 2 d^2 j^2 \pi^2 \sin^2{\varphi}\cos{\left(2dj\pi\cos{\varphi}\right)}
\cos{\left(\textstyle{\frac{2dk\pi\sin{\varphi}}{\alpha}}\right)} \nonumber\\
& & + d j \pi\cos{\varphi} \sin{\left(2dj\pi\cos{\varphi}\right)}
\cos{\left(\textstyle{\frac{2dk\pi\sin{\varphi}}{\alpha}}\right)} \nonumber\\
& & + \frac{dk\pi}{\alpha} \sin{\varphi} \cos{\left(2dj\pi\cos{\varphi}\right)}
\sin{\left(\textstyle{\frac{2dk\pi\sin{\varphi}}{\alpha}}\right)} \nonumber\\
& & - \frac{4d^2jk\pi^2}{\alpha} \cos{\varphi} \sin{\varphi}
\sin{\left(2dj\pi\cos{\varphi}\right)}
\sin{\left(\textstyle{\frac{2dk\pi\sin{\varphi}}{\alpha}}\right)} \Bigg]
\nonumber \\ & & \times
z_{3/2}\left(\textstyle{j^2\alpha+\frac{k^2}{\alpha}}\right) \nonumber\\
& & + \frac{d}{2} \sum_{j,k=-\infty}^\infty \left( \frac{j\cos{\varphi}}{\alpha}
+ k \sin{\varphi}\right) \nonumber \\ & & \times
z_{5/2}\left[\textstyle{\frac{(j+d\cos{\varphi})^2}{\alpha}+\alpha
\left(k+\frac{d\sin{\varphi}}{\alpha}\right)^2}\right] \nonumber \\
& & + d^2 \sum_{j,k=-\infty}^\infty \left(k\cos{\varphi} - 
\frac{j\sin{\varphi}}{\alpha}\right)^2 \nonumber \\ & & \times
z_{7/2}\left[\textstyle{\frac{(j+d\cos{\varphi})^2}{\alpha}+\alpha
\left(k+\frac{d\sin{\varphi}}{\alpha}\right)^2}\right] \nonumber \\
& & + \sum_{j=1}^\infty \Bigg\{ \big[ - 2 d^2 j^2 \pi^2
\cos{\left(j\pi+2dj\pi\cos{\varphi}\right)} \sin^2{\varphi} \nonumber \\
& & + d j \pi \cos{\varphi} \sin{\left(j\pi+2dj\pi\cos{\varphi}\right)} \big] 
z_{3/2}(j^2\alpha) \nonumber \\
& & + \frac{1}{\alpha^2} \Bigg[ -2 d^2 j^2 \pi^2 \cos^2{\varphi}
\cos{\left(\textstyle{j\pi+\frac{2dj\pi\sin{\varphi}}{\alpha}}\right)}
\nonumber\\ & & + \alpha d j \pi \sin{\varphi}
\sin{\left(\textstyle{j\pi+\frac{2dj\pi\sin{\varphi}}{\alpha}}\right)} 
\Bigg] z_{3/2}\left(\textstyle{\frac{j^2}{\alpha}}\right) \Bigg\} \nonumber \\
& & + 2 \sum_{j,k=1}^\infty \Bigg[
-\frac{2d^2k^2\pi^2\cos^2{\varphi}}{\alpha^2}
\cos{\left(j\pi+2dj\pi\cos{\varphi}\right)} \nonumber \\ & & \times
\cos{\left(\textstyle{k\pi+\frac{2dk\pi\sin{\varphi}}{\alpha}}\right)} 
\nonumber\\
& & - 2 d^2 j^2 \pi^2 \sin^2{\varphi} 
\cos{\left(j\pi+2dj\pi\cos{\varphi}\right)}
\cos{\left(\textstyle{k\pi+\frac{2dk\pi\sin{\varphi}}{\alpha}}\right)}
\nonumber\\ & & + d j \pi \cos{\varphi}
\sin{\left(j\pi+2dj\pi\cos{\varphi}\right)}
\cos{\left(\textstyle{k\pi+\frac{2dk\pi\sin{\varphi}}{\alpha}}\right)}
\nonumber\\ & & + \frac{dk\pi}{\alpha} \sin{\varphi}
\cos{\left(j\pi+2dj\pi\cos{\varphi}\right)}
\sin{\left(\textstyle{k\pi+\frac{2dk\pi\sin{\varphi}}{\alpha}}\right)}
\nonumber\\ & & - \frac{4d^2\pi^2jk}{\alpha} \cos{\varphi} \sin{\varphi}
\sin{\left(j\pi+2dj\pi\cos{\varphi}\right)} \nonumber \\ & & \times 
\sin{\left(\textstyle{k\pi+\frac{2dk\pi\sin{\varphi}}{\alpha}}\right)}
\Bigg] z_{3/2}\left(\textstyle{j^2\alpha+\frac{k^2}{\alpha}}\right) \nonumber \\
& & + \frac{d}{2} \sum_{j,k=-\infty}^\infty \left[
\frac{(1+2j)\cos{\varphi}}{2\alpha} + \left( k+\frac{1}{2}\right)
\sin{\varphi}\right] \nonumber \\ & & \times 
z_{5/2}\left[\textstyle{\frac{\left(j+1/2+d\cos{\varphi}\right)^2}{\alpha}
+\alpha \left(k+\frac{1}{2}+\frac{d\sin{\varphi}}{\alpha}\right)^2}\right]
\nonumber \\ & & + \frac{d^2}{4} \sum_{j,k=-\infty}^\infty 
\left[ - \frac{(1+2j)\sin{\varphi}}{\alpha} + (1+2k)\cos{\varphi}\right]^2
\nonumber \\ & & \times 
z_{7/2}\left[\textstyle{\frac{\left(j+1/2+d\cos{\varphi}\right)^2}{\alpha}
+\alpha\left(k+\frac{1}{2}+\frac{d\sin{\varphi}}{\alpha}\right)^2}\right] .
\end{eqnarray}

\section{The small-$d$ expansion of the energy} \label{C}
Using the expansion of the Misra function (\ref{z32ex}), the small-$d$ 
expansion of the energy $E_0(\varphi,d)$ (\ref{e0}) takes the form
(\ref{e0exd}), where
\begin{eqnarray}
f_2(\varphi) & = & - 16 \pi^2 \sum_{j=1}^\infty 
\left[j^2 \cos^2{\varphi}\, z_{3/2}(j^2\alpha) +
\frac{j^2\sin^2{\varphi}}{\alpha^2}
z_{3/2}\left(\textstyle{\frac{j^2}{\alpha}}\right) \right] \nonumber \\
& & - 32 \pi^2 \sum_{j,k=1}^\infty \left[j^2\cos^2{\varphi}
+ \frac{k^2\sin^2{\varphi}}{\alpha^2} \right]
z_{3/2}\left(\textstyle{j^2\alpha+\frac{k^2}{\alpha}}\right) \nonumber \\
& & - \frac{4}{\alpha} {\sum_{j,k=-\infty\atop (j,k)\ne (0,0)}^{\infty}} 
z_{5/2}\left(\textstyle{\frac{j^2}{\alpha}+k^2\alpha}\right) \nonumber \\
& & + 8 \sum_{j,k=-\infty\atop (j,k)\ne (0,0)}^\infty 
\left(\frac{j\cos{\varphi}}{\alpha}+k\sin{\varphi}\right)^2
z_{7/2}\left(\textstyle{\frac{j^2}{\alpha}+k^2\alpha}\right) \label{f2}
\end{eqnarray}
and
\begin{eqnarray} 
f_4(\varphi) & = & \frac{64}{3} \pi^4 \sum_{j=1}^\infty \left[
j^4 \cos^4{\varphi}\, z_{3/2}(j^2\alpha) + \frac{j^4\sin^4{\varphi}}{\alpha^4}
z_{3/2}\left(\textstyle{\frac{j^2}{\alpha}}\right)\right] \nonumber\\
& & + \frac{128}{3} \pi^4 \sum_{j,k=1}^\infty \bigg[ j^4 \cos^4{\varphi}
+\frac{6}{\alpha^2} j^4 \cos^2{\varphi} \sin^2{\varphi} \nonumber \\
& & + \frac{1}{\alpha^4} j^4 \sin^4{\varphi} \bigg]
z_{3/2}\left(\textstyle{j^2\alpha+\frac{k^2}{\alpha}}\right) 
+ \frac{8}{\alpha^2} \sum_{j,k=-\infty\atop (j,k)\ne (0,0)}^\infty 
z_{7/2}\left(\textstyle{\frac{j^2}{\alpha}+k^2\alpha}\right) \nonumber\\
& & + \frac{32}{\alpha^3} \sum_{j,k=-\infty\atop (j,k)\ne (0,0)}^\infty 
\left(j\cos{\varphi} +\alpha k\sin{\varphi}\right)^2
z_{9/2}\left(\textstyle{\frac{j^2}{\alpha}+k^2\alpha}\right) \nonumber \\
& & + \frac{32}{3} \sum_{j,k=-\infty\atop (j,k)\ne (0,0)}^\infty
\left(\frac{j\cos{\varphi}}{\alpha} + k \sin{\varphi}\right)^4
z_{11/2}\left(\textstyle{\frac{j^2}{\alpha}+k^2\alpha}\right) . \label{f4}
\end{eqnarray}

Now we sketch the derivation of the form of the functions $f_2$ and $f_4$
for the specific case of the square lattice ($\alpha=1$). 
First we shall show that $f_2(\varphi)$ is constant.
The first sum in (\ref{f2}) becomes $\varphi$-independent because of 
the equality $\cos^2{(\varphi)} + \sin^2{(\varphi)} = 1$,
the second one after combining symmetrically the $(j,k)$ and $(k,j)$ summands.
In the last sum we combine the $j,k$ and $k,-j$ summands to get 
$(j\cos{\varphi}+k\sin{\varphi})^2+(k\cos{\varphi}-j\sin{\varphi})^2 = j^2+k^2$.
As concerns the function $f_4(\varphi)$, we shall show that it is a linear 
function of $\cos{(4\varphi)}$.
We combine again $j,k$ and $k,\pm j$ summands, and vice versa, and apply 
relations like
$4(\cos^4{(\varphi)}+\sin^4{(\varphi)}) = 3+\cos{(4\varphi})$, 
$8\cos^2{(\varphi)}\sin^2{(\varphi)})=1-\cos{(4\varphi)}$
and
\begin{eqnarray} 
& & (j\cos{\varphi} +k\sin{\varphi})^4 + (j\cos{\varphi} -k\sin{\varphi})^4
+(k\cos{\varphi} +j\sin{\varphi})^4 \nonumber\\
& & + (k\cos{\varphi} -j\sin{\varphi})^4 = \frac{3}{2}(j^2+k^2)^2
+\left(\frac{j^4+k^4}{2}-3j^2k^2\right) \cos{(4\varphi)} . \label{cos4phi}
\end{eqnarray}
Knowing that $f_4(\varphi)$ is composed of two terms, the absolute one
and the one linear in $\cos{(4\varphi)}$, the two prefactors in 
relation (\ref{f4exact}) are determined uniquely.

\end{document}